\documentclass[twocolumn]{aastex631}

\usepackage{float}

\usepackage{threeparttable}

\usepackage{comment}

\usepackage{natbib}
\usepackage{multirow}
\usepackage{color}
\usepackage{bm}
\usepackage[normalem]{ulem}
 \usepackage{amsmath} 
\usepackage {diagbox}
\usepackage{ltablex}
\usepackage{longtable}

\newcommand{\Fefs}{$^{56}$Fe}
\newcommand{\Cofs}{$^{56}$Co}
\newcommand{\Nifs}{$^{56}$Ni}

\def\gsim{\mathrel{\rlap{\lower 4pt \hbox{\hskip 1pt $\sim$}}\raise 1pt \hbox {$>$}}}
\def\lsim{\mathrel{\rlap{\lower 4pt \hbox{\hskip 1pt $\sim$}}\raise 1pt \hbox {$<$}}}

\newcommand{\deltat}{$\Delta t_{\rm 0.5mag}$}
\newcommand{\Mpeak}{$M_{\rm peak}$}
\newcommand{\Mej}{$M_{\rm ej}$}
\newcommand{\Msun}{$M_\odot$}
\newcommand{\MNi}{$M_{\rm Ni}$}
\newcommand{\vej}{$v_{\rm ej}$}
\newcommand{\tgrow}{$t_{\rm grow}$}

\usepackage{ulem}

\newcommand{\ltsim}{\protect\raisebox{-0.8ex}{$\:\stackrel{\textstyle <}{\sim}\:$}}
\newcommand{\gtsim}{\protect\raisebox{-0.8ex}{$\:\stackrel{\textstyle >}{\sim}\:$}}

\newcommand{\pd}[2]{\frac{\partial #1}{\partial #2}}
\newcommand{\ld}[2]{\frac{{ D} #1}{{ D} #2}}

\shorttitle{Constraints on explosion timescale from \Nifs ~ mass}
\shortauthors{S.Saito et al.}

\begin{document}

\title{Constraints on Explosion Timescale of Core-Collapse Supernovae \\
Based on Systematic Analysis of Light Curves}

\correspondingauthor{Sei Saito}
\email{s.saito@astr.tohoku.ac.jp}

\author{Sei Saito}
\affiliation{Astronomical Institute, Tohoku University, Sendai 980-8578, Japan}

\author{Masaomi Tanaka}
\affiliation{Astronomical Institute, Tohoku University, Sendai 980-8578, Japan}
\affiliation{Division for the Establishment of Frontier Sciences, Organization for Advanced Studies, Tohoku University, Sendai 980-8577, Japan}

\author[0000-0003-4876-5996]{Ryo Sawada}
\affiliation{Department of Earth Science and Astronomy, Graduate School of Arts and Sciences, The University of Tokyo, Tokyo 153-8902, Japan}

\author[0000-0003-1169-1954]{Takashi J. Moriya}
\affiliation{Division of Science, National Astronomical Observatory of Japan, National Institutes of Natural Sciences, 2-21-1 Osawa, Mitaka, Tokyo 181-8588, Japan}
\affiliation{School of Physics and Astronomy, Faculty of Science, Monash University, Clayton, Victoria 2800, Australia}

\begin{abstract}

Explosion mechanism of core-collapse supernovae is not fully understood yet.
In this work, we give constraints on the explosion timescale based on \Nifs ~ synthesized by supernova explosions.
First, we systematically analyze multi-band light curves of 82 stripped-envelope supernovae (SESNe)
to obtain bolometric light curves,
which is among the largest samples of the bolometric light curves of SESNe
derived from the multi-band spectral energy distribution.
We measure the decline timescale and the peak luminosity of the light curves
and estimate the ejecta mass (\Mej) and \Nifs ~ mass (\MNi) to connect the observed properties with the explosion physics.
We then carry out one-dimensional hydrodynamics and nucleosynthesis calculations,
varying the progenitor mass and the explosion timescale.
From the calculations, we show that the maximum \Nifs ~ mass that \Nifs-powered SNe can reach
is expressed as $M_{\rm Ni} \ltsim 0.2 \ M_{\rm ej}$.
Comparing the results from the observations and the calculations,
we show that the explosion timescale shorter than 0.3 sec explains the synthesized \Nifs~ mass of the majority of the SESNe.

 \end{abstract}


\section{Introduction}
\label{sec:int}

Massive stars whose initial masses are \gtsim 8 \Msun ~ explode
at the end of their lives as core-collapse supernovae (CCSNe).
After the core collapses, 
the central density increases, 
which eventually stops the core contraction.
This causes a bounce of the core, generating a shock wave.
If the shock wave propagates to the surface of the star,
the supernova (SN) explosion is considered to be successful.
However, the explosion mechanism of CCSNe is not fully understood yet.
The propagation of the shock wave is halted
by endothermal reactions due to photodisintegration of iron
and by ram pressure of falling materials from the outer part.
Successful explosions thus require a mechanism to revive the shock wave.
The most plausible mechanism is a neutrino-driven explosion mechanism \citep{Arnett1966, Colgate1966, Bethe1985}.
In this mechanism, 
part of the neutrinos produced in the dense core are absorbed by materials behind the shock wave,
reviving the shock wave.
As a result, the SN explosion driven by neutrinos can occur.

It has been known that in multi-dimensional ab-initio calculations with the neutrino-driven explosion mechanism
\citep[e.g., ][]{Janka2012, Takiwaki2014, Lentz2015, Melson2015a, Melson2015b, Muller2017, Muller2019,
Ott2018, Summa2018, Burrows2019, Burrows2020, Stockinger2020},
the kinetic energy only reaches $E_{\rm k} \sim 10^{50}$ erg,
while a typical energy estimated from observations is $E_{\rm k} \sim 10^{51}$ erg
\citep[e.g.,][]{Hamuy2003, Drout2011, Cano2013, Pejcha2015, Taddia2015, Taddia2018, Lyman2016, Barbarino2021}.
Some of recent simulations have been reported to reach $E_{\rm k} = 10^{51}$ erg \citep[e.g.,][]{Burrows2021, Bollig2021},
and the problem in the explosion energy is gradually being solved. 
It should be noted, however, that many simulations do not reproduce energetic explosion. 
Moreover, the timescales to reach the sufficient kinetic energy in those calculations are $\sim$ 1 sec or even longer.

Such a slow explosion tends to result in a relatively small production of \Nifs ~ \citep[$\sim 0.03$ \Msun; e.g., ][]{Muller2017},
which is a radioactive nucleus powering SNe by the radioactive decay through \Cofs ~ to \Fefs ~ \citep{Arnett1966}.
In models with calibrated neutrino luminosity to reproduce SN explosions, 
the synthesized \Nifs ~ mass can be $\ltsim 0.1$ \Msun ~ \citep[e.g.,][]{Ugliano2012, Ertl2020, Woosley2021},
which is higher than that from ab-initio calculations.

The \Nifs ~ mass has been estimated from observations.
For stripped-envelope SNe (SESNe; Type IIb, Ib, Ic, and Ic BL (standing for broad line)) without hydrogen-rich envelopes,
the \Nifs ~ mass is often estimated from the peak of their light curves
based on a semi-analytic model \citep[Arnett's rule;][]{Arnett1982}.
The range of the mass is $\sim 0.0015 \ M_\odot \ - \sim 1$ \Msun ~
with the median value of $\sim$ 0.15 \Msun ~ \citep{Anderson2019, Meza2020}.
On the other hand, for Type II SNe with hydrogen-rich envelopes,
the \Nifs ~ mass is estimated from the tail of their light curves \citep[from the \Cofs ~ decay, e.g.,][]{Woosley1988}.
The range of the mass is $\sim 0.001 \ M_\odot \ - \sim 0.3$ \Msun ~
with the median value of $\sim$ 0.03 \Msun ~ \citep{Anderson2019, Meza2020}.
It should be noted, however, that lack of the \Nifs ~ mass in the lower side of SESNe can be explained by observation bias
\citep{Ouchi2021}.

These observational facts imply that the range of the \Nifs ~ mass estimated from observations cannot easily be explained by
ab-initio calculations nor models with calibrated-neutrino luminosity.
For example, \citet{Sollerman2021} compared the peak luminosities of many observed SESNe in optical bands
with those expected from the models with calibrated neutrino luminosity \citep{Woosley2021}.
They have shown that only $\sim 60 ~ \%$ of the luminosities of the observed SESNe can be reproduced
by the model.

To understand the necessary condition of CCSNe,
several works have been done based on “artificial-explosion" calculations \citep[e.g., ][]{Maeda2009, Suwa2015, Suwa2019}.
Those studies have pointed out that the synthesized \Nifs ~ mass increases with the decrease of the explosion timescale.

While the explosion timescale of \gtsim 1 sec (“slow” explosions) has been predicted by the current ab-initio calculations,
\citet{Sawada2019} has shown that the shorter explosion timescale of $\ltsim 0.25$ sec (“rapid” explosions) 
is required to reproduce the typical mass of \Nifs.
However, they did not compare their results with concrete observational data 
and only showed a general trend in comparison with some progenitors of typical SNe.
Here, various progenitors, together with various explosion timescales,
should be considered to estimate \Nifs ~ mass
because the production of \Nifs ~ depends not only on the explosion timescale but also on progenitor stars
\citep[e.g.,][]{Woosley1995, Sukhbold2016}.

In this work, we compare \Nifs ~ masses estimated from observations and from calculations varying progenitor stars.
We perform the systematic analysis of observed light curves of SESNe
including Type I superluminous supernovae (Type I SLSNe or SLSNe-I; hydrogen-poor SLSNe) 
whose peak magnitudes are $\ltsim -21$ mag \citep[e.g.,][]{GalYam2012, Moriya2018a, Nicholl2021}.
Although the high luminosity of SLSNe is difficult to be explained by \Nifs,
it is not clear up to what luminosity radioactive decay of \Nifs ~ can serve as a power source.

This paper is organized as follows.
Section \ref{sec:data} describes methods to systematically analyze multi-band light curves of SESNe
to estimate the synthesized the \Nifs ~ mass with the maximum sample number to date.
Section \ref{sec:res} shows the results of the systematic analysis of the light curves.
In Section \ref{sec:dis}, we carry out one-dimensional hydrodynamics and nucleosynthesis calculations
to investigate the synthesized \Nifs ~ mass.
Furthermore, we discuss the maximum luminosity of \Nifs-powered SNe and the explosion timescale of CCSNe.
Finally, the conclusions are given in Section \ref{sec:sum}.

\section{Data selection and analysis}
\label{sec:data}

\begin{figure}[ht]
\centering
\includegraphics[width=0.7\linewidth]{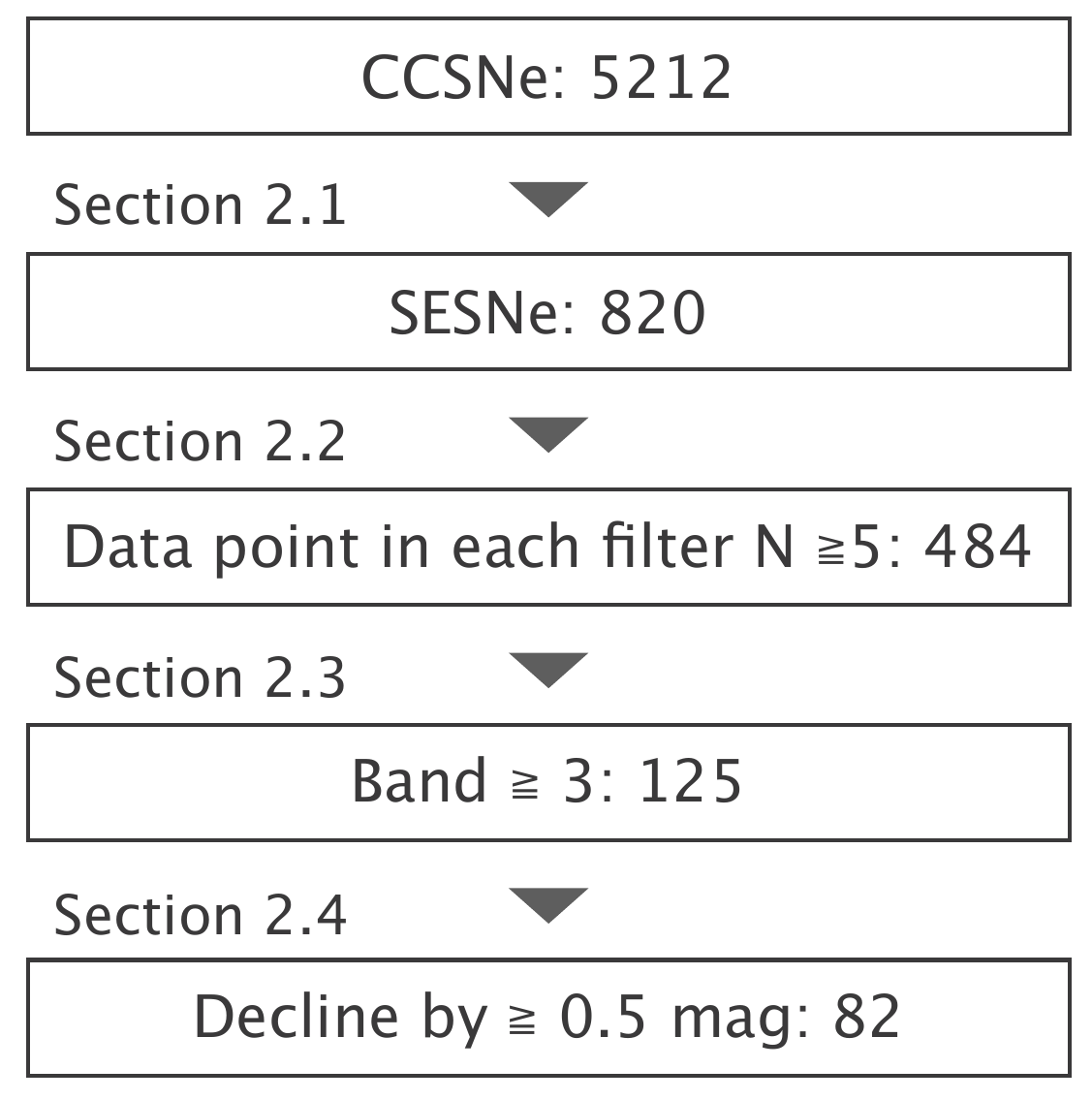}
\caption{
The procedure to select our sample with the section number and the number of the sample.}
\label{fig:select}
\end{figure}

In this section, we describe observational data used in our analysis and methods of the analysis.
We derive bolometric light curves of 82 SESNe from the multi-band observational data
and measure decline timescales and peak magnitudes from the bolometric light curves.
The sample selection and the procedure of the analysis are shown in Figure \ref{fig:select}.

\subsection{Types of supernovae}
\label{data:sample}

We used photometric data of SNe from a SN database, Open Supernova Catalog \citep{guillochon2017jan}.
The catalog has 59603 objects as of 2020 Mar 13, including 5212 CCSNe.
We selected SESNe including SLSNe-I from the sample, finding 820 SESNe.
The reason why we only selected SESNe is that their \Nifs ~ mass can be estimated from the peak luminosity.
Furthermore, we can estimate the ejecta mass from the light curve,
which provides the connection with the progenitor stars.

\subsection{Interpolation of photometric data}
\label{data:gpr}

We smoothly connect the data points of the SESNe selected in the previous section to interpolate the data points,
which are inevitably caused by intervals of observations.
To ensure the reliable interpolation,
we only use the data with the observational points $N \geq 5$ for each filter.
After this criterion, 484 SNe remain.

\begin{figure*}[ht]
\begin{center}
  \begin{tabular}{c}
    \begin{minipage}{0.5\hsize}
      \begin{center}
       \includegraphics[width= 0.97\linewidth, bb=7 7 460 580]{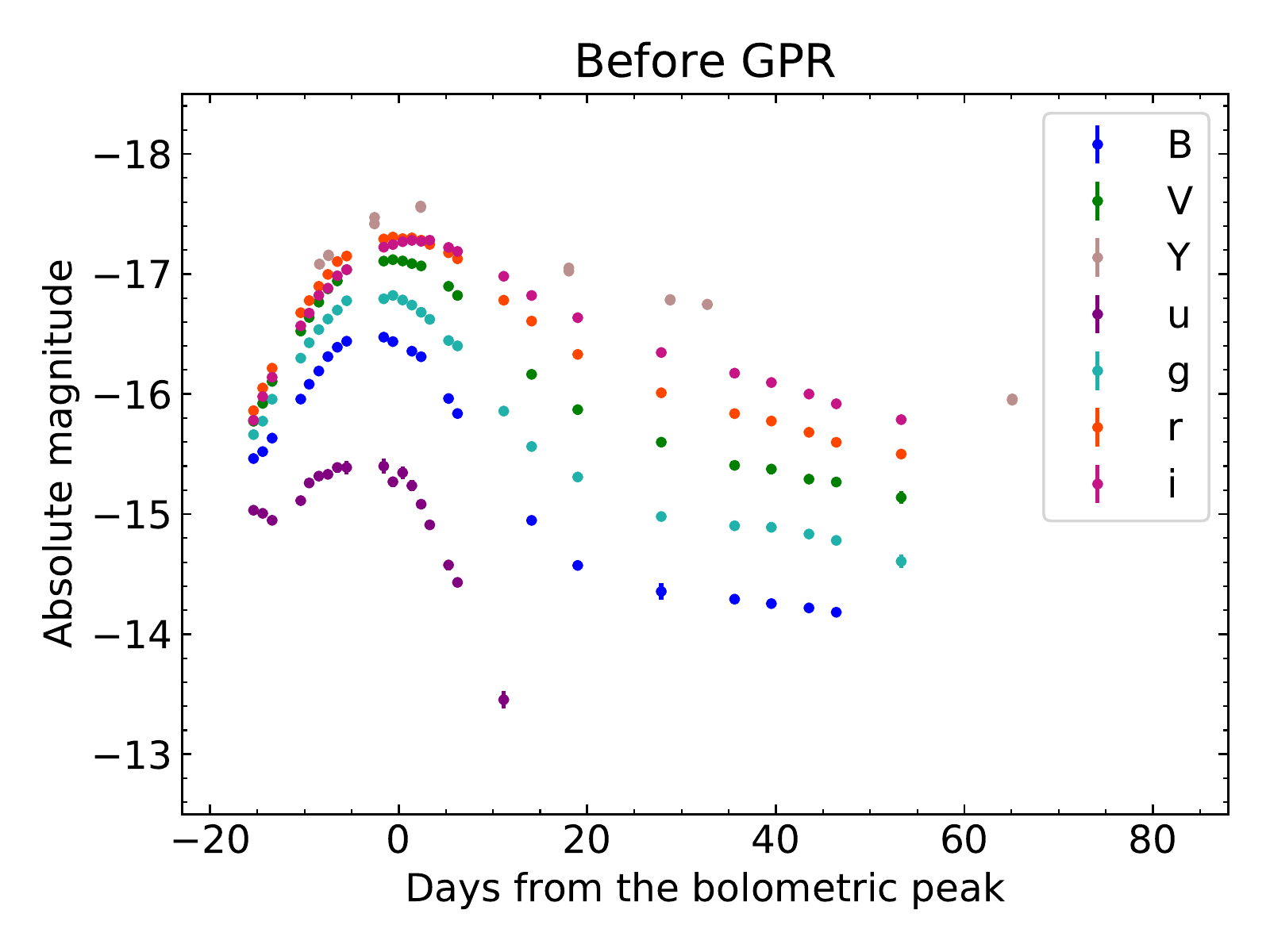}
        \end{center}
    \end{minipage}    
     \begin{minipage}{0.5\hsize}
      \begin{center}
       \includegraphics[width= 0.97\linewidth, bb=7 7 460 580]{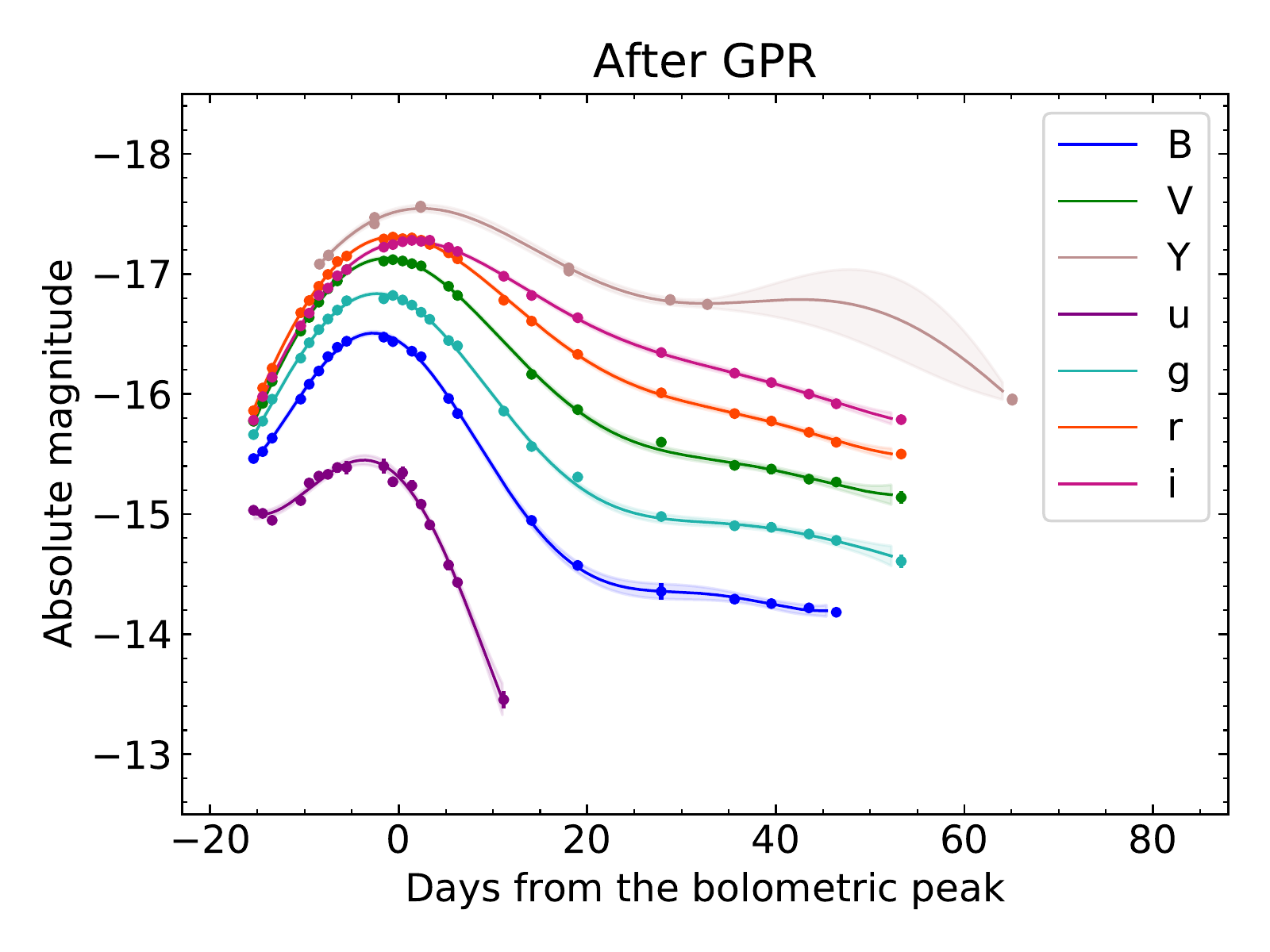}
       \end{center}
    \end{minipage}
  \end{tabular}  
\caption{An example of multi-band light curves before (the left panel) and after (the right panel) the GPR (Type IIb SN 2004ex).
The different colors show different bands as shown in the legend.}
  \label{fig:GPR}
  \end{center}
\end{figure*}

For the interpolation of data points,
we performed Gaussian Process Regressions (GPRs),
which can express complex nonlinear functions by superposition of infinite number of functions.
The kernel of the GPR used for fitting is \textit{constant} $\times$ Radial Basis Functions.
The kernel length parameter is fitted to obtain the best match with the data.
Figure \ref{fig:GPR} shows a comparison of data before the GPR and after the GPR,
demonstrating that the observational data points are smoothly fitted with the GPR.

Here, we discuss some light curves which we manually eliminated from our data analysis.
Some light curves
cannot be fitted well by the GPR method
due to the sparseness of data points.
The eliminated light curves of SNe and bands are following.
PS16yj: $UVM2$;
PS1-12sk: $UVW1, UVW2, UVM2$;
PTF09cnd: $UVW1, UVW2, UVM2$;
PTF12dam: $UVW1, UVW2, U, i$;
SN 2004dk: $I$;
SN 2005az: $R, r$;
SN 2007Y: $U$;
SN 2009jf: $u, i$;
SN 2009mg: $UVW1$;
SN 2010bh: $B$;
SN 2011kg: $UVW1, UVW2, u, U$;
SN 2013hy: $g$;
SN 2014ad: $UVW1, UVW2, UVM2$;
SN 2015bn: $V, i, z$;
SN 2016coi: $U, I$;
SN 2017ein: $I$;
iPTF13bvn: $UVW2, U$.

\subsection{Bolometric light curve}
\label{data3}

We here derive bolometric light curves using multi-band light curves obtained above.
First, we correct the dust extinction.
Then, we take the luminosity distances $d_{\rm L}$ to obtain the absolute magnitudes.
After that, we perform blackbody fitting to derive bolometric light curves.
In our analysis, the Vega magnitudes have been converted to the AB magnitudes based on \citet{Blanton2007}.

The extinction of the Milky Way was corrected based on the dust map estimated in \citet{Schlegel1998}.
On the other hand, the extinction of host galaxies of SNe was not corrected
since the degree of host extinction is uncertain.
It is noted that although the typical value of host extinction is 
$A_V \sim 0.6$ for Type IIb and Ib SNe and $A_V \sim 1$ for Type Ic SNe \citep{Stritzinger2018}
or $A_V \sim 0.9$ for SESNe \citep{Zheng2022},
ignoring the host extinction does not affect on the conclusion of this work.
This is because we will use the upper limit of the luminosity in our discussion (Section \ref{dis2}).

\begin{figure}[ht]
\centering
\includegraphics[width=\linewidth]{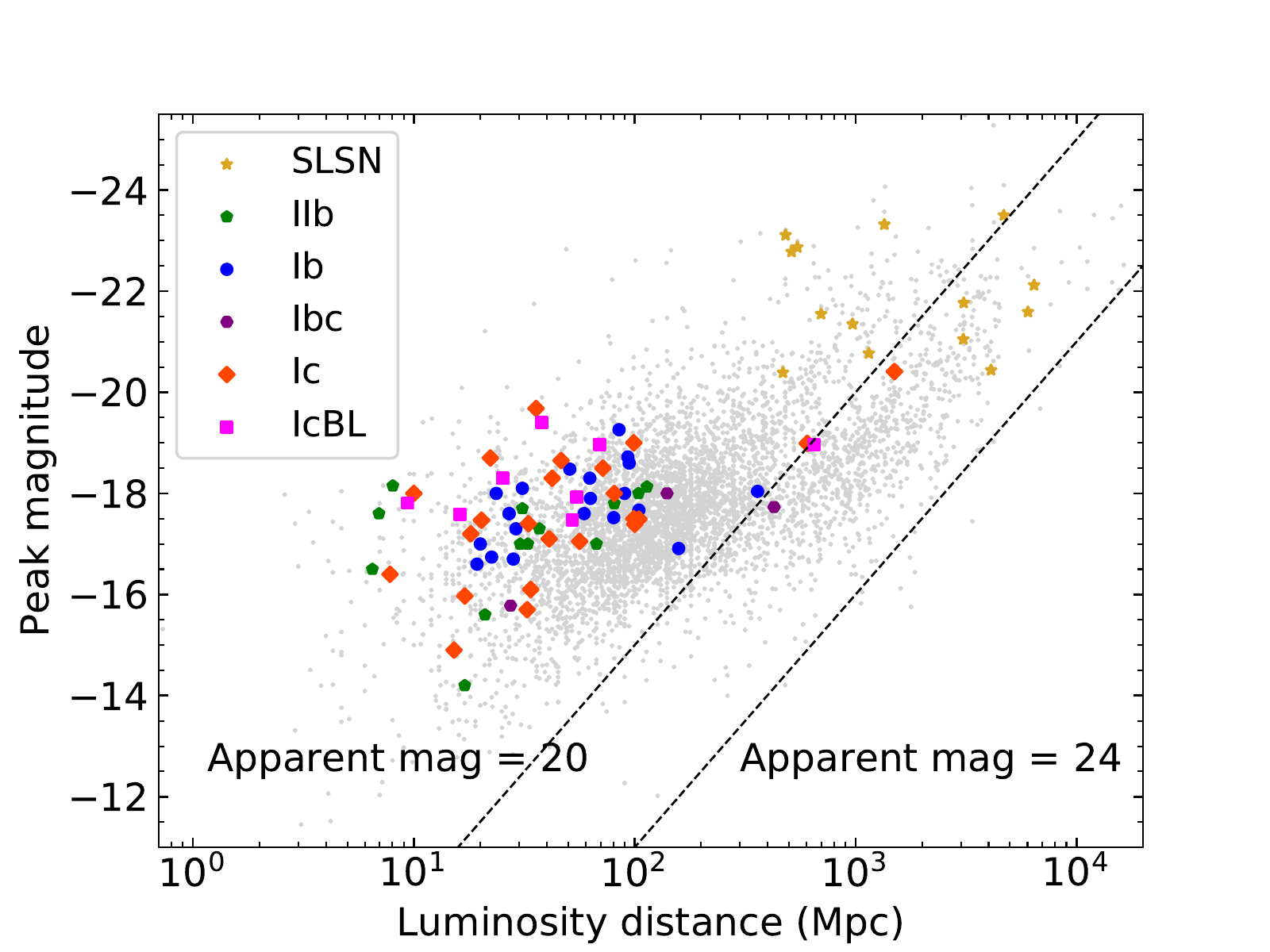}
\caption{
A relation between luminosity distances and peak absolute magnitudes
in the band that shows the maximum brightness among all the bands used for each object.
The gray points show all the CCSNe in Open Supernova Catalog
, and the colored points show our final samples. 
The different colors represent different types of SNe as shown in the legend.
Note that the peak magnitudes are different from the bolometric magnitudes in Figure \ref{fig:deltat_Mpeak}.}
\label{fig:sample}
\end{figure}

To obtain the absolute magnitudes, we took the luminosity distances $d_{\rm L}$ from Open Supernova Catalog.
For the distances for which the method of the estimate is uncertain,
we instead used the distances based on
the Cepheid measurement (\ltsim 10 Mpc),
the Tully Fisher relation \citep[$\sim$ 10 $- \sim $ 20 Mpc][]{Tully1977}, 
and the redshift with a correction from Galactic infall model (\gtsim 20 Mpc).
Figure \ref{fig:sample} represents a relation between luminosity distances and peak absolute magnitudes
in the band that shows the maximum brightness among all the bands used for each object.
The luminosity distances $d_{\rm L}$ are mainly in the range of $10 \ltsim d_{\rm L} \ltsim 1000$ Mpc
and the peak magnitudes range from $-14$ to $-24$ mag.
More luminous objects tend to be observed at larger distances due to the observational bias.

\begin{figure}[ht]
\centering
\includegraphics[width=\linewidth]{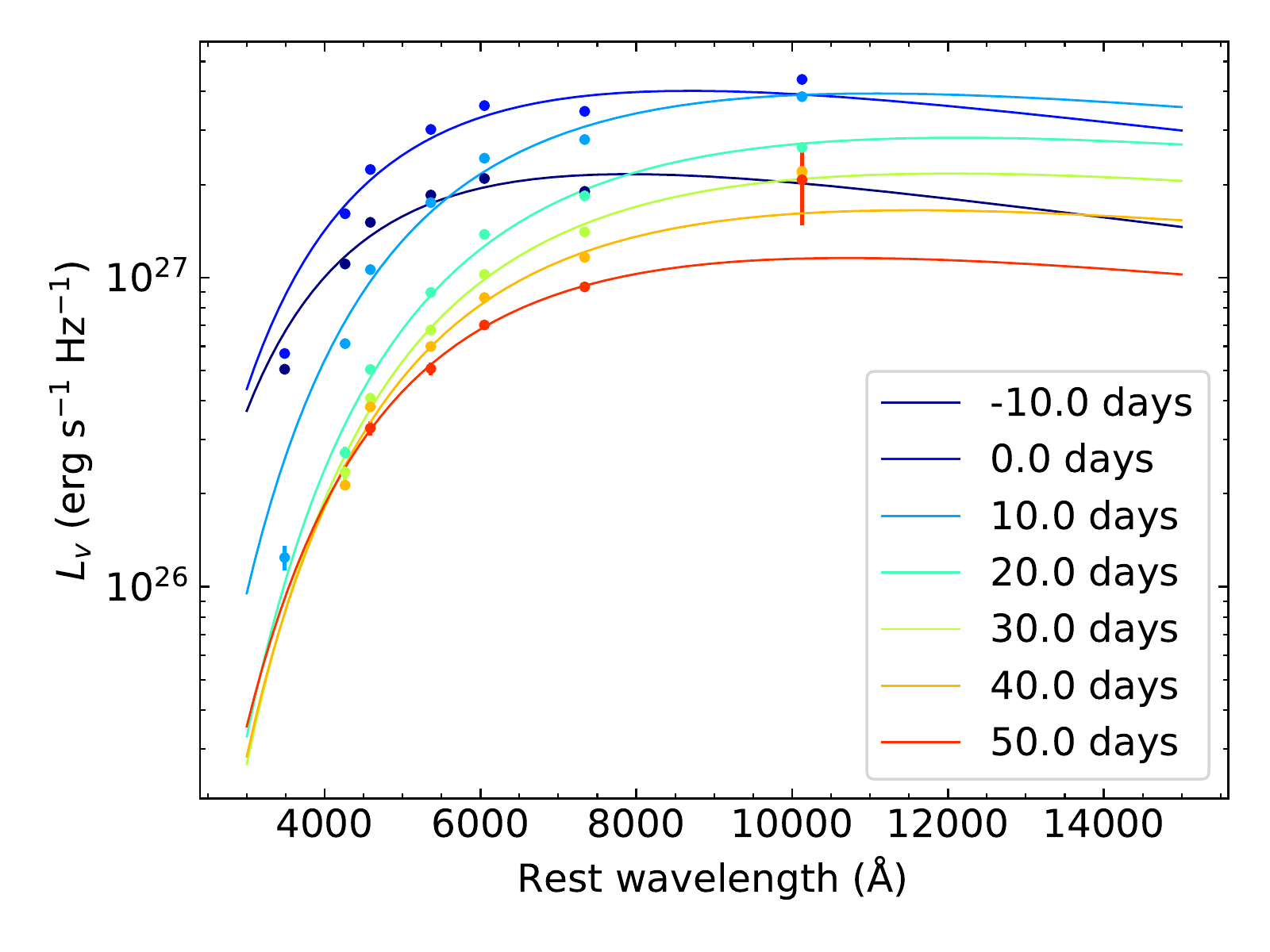}
\caption{An example of time evolution of SEDs (SN 2004ex).
The different colors show different epochs in the rest frame from the bolometric peak luminosity, as shown in the legend.}
\label{fig:SED}
\end{figure}

To derive bolometric luminosities, 
we performed the blackbody fitting to the spectral energy distributions (SEDs) estimated with the methods in Section \ref{data:gpr}
if a SN was observed with $\geq 3$ bands.
The spectra can be approximated with a blackbody \citep[e.g., ][]{Modjaz2009}
because SNe in photospheric phases mainly have thermal radiation
typically until $\sim 50$ days after the explosion.
Figure \ref{fig:SED} shows an example of SEDs with the time evolution.
The SEDs are approximated well with a blackbody radiation until 50 days after the explosion.
By integrating the fitted flux over the entire wavelength,
we have obtained bolometric light curves of 125 SNe.

\subsection{Measurement of decline timescale and peak magnitude}
\label{data4}

From the bolometric light curves constructed in Section \ref{data3},
we measure peak bolometric absolute magnitudes $M_{\rm peak}$ and decline timescales \deltat.
Here, \deltat ~ is defined as the time it takes for the light curve to decline by 0.5 mag from the peak luminosity.
which is the time for the light curve to decline by 0.5 mag from the peak.
We finally measured the peak absolute magnitudes and the decline timescales of 82 SNe.
The sample consists of 14 SLSNe-I, 13 Type IIb, 21 Type Ib, 24 Type Ic, 6 Type Ic-BL and 4 Type Ibc.

\section{Results}
\label{sec:res}

\subsection{\deltat ~ vs. \Mpeak}
\label{res:mag}

\begin{figure*}[ht]
\centering
\includegraphics[width= 0.9\linewidth]{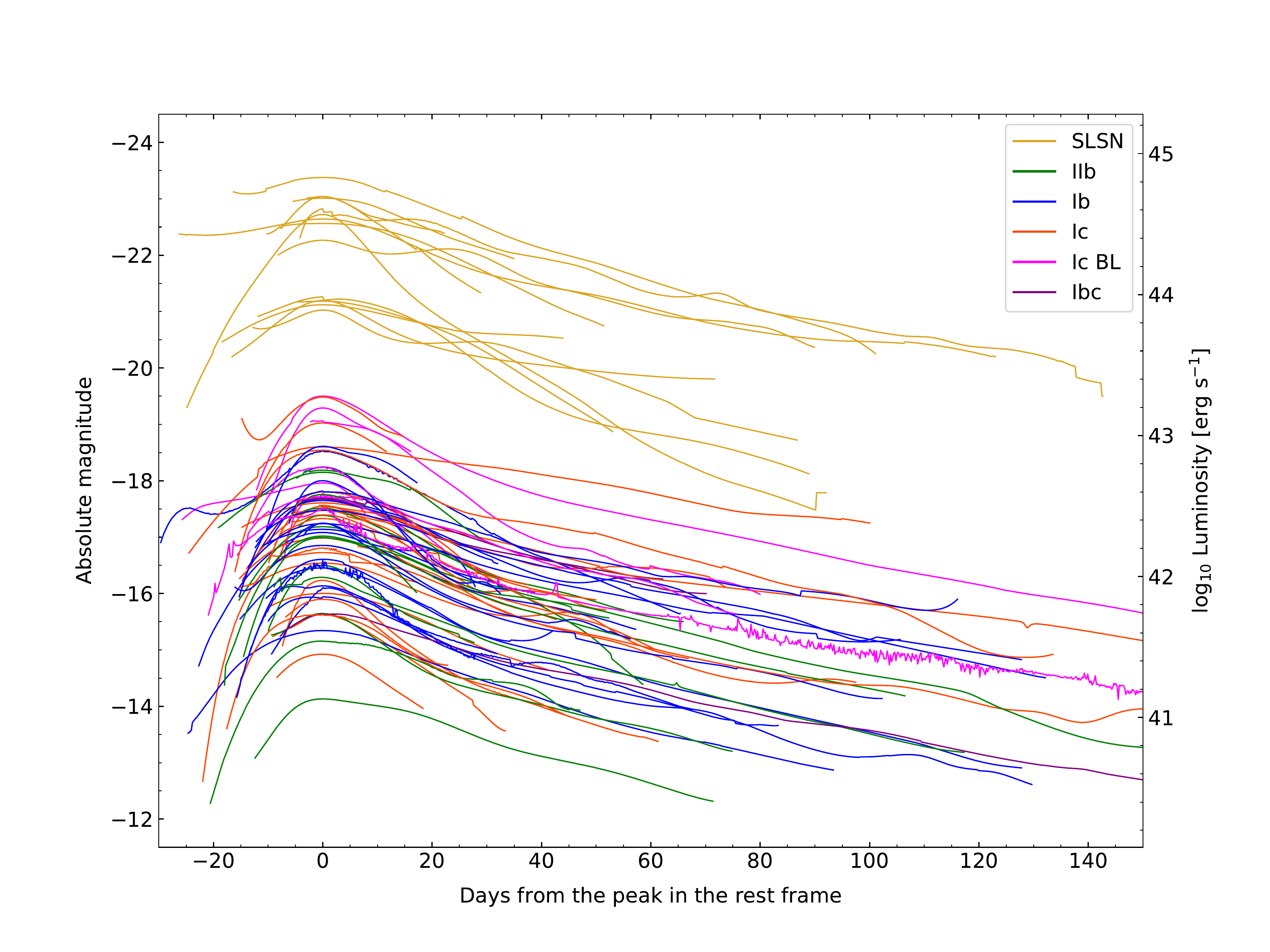}
\caption{Bolometric light curves of our samples.
The different colors show different types of SNe as shown in the legend.
}
\label{fig:bol}
\end{figure*}

\begin{figure*}[ht]
\centering
\includegraphics[width= 0.7\linewidth]{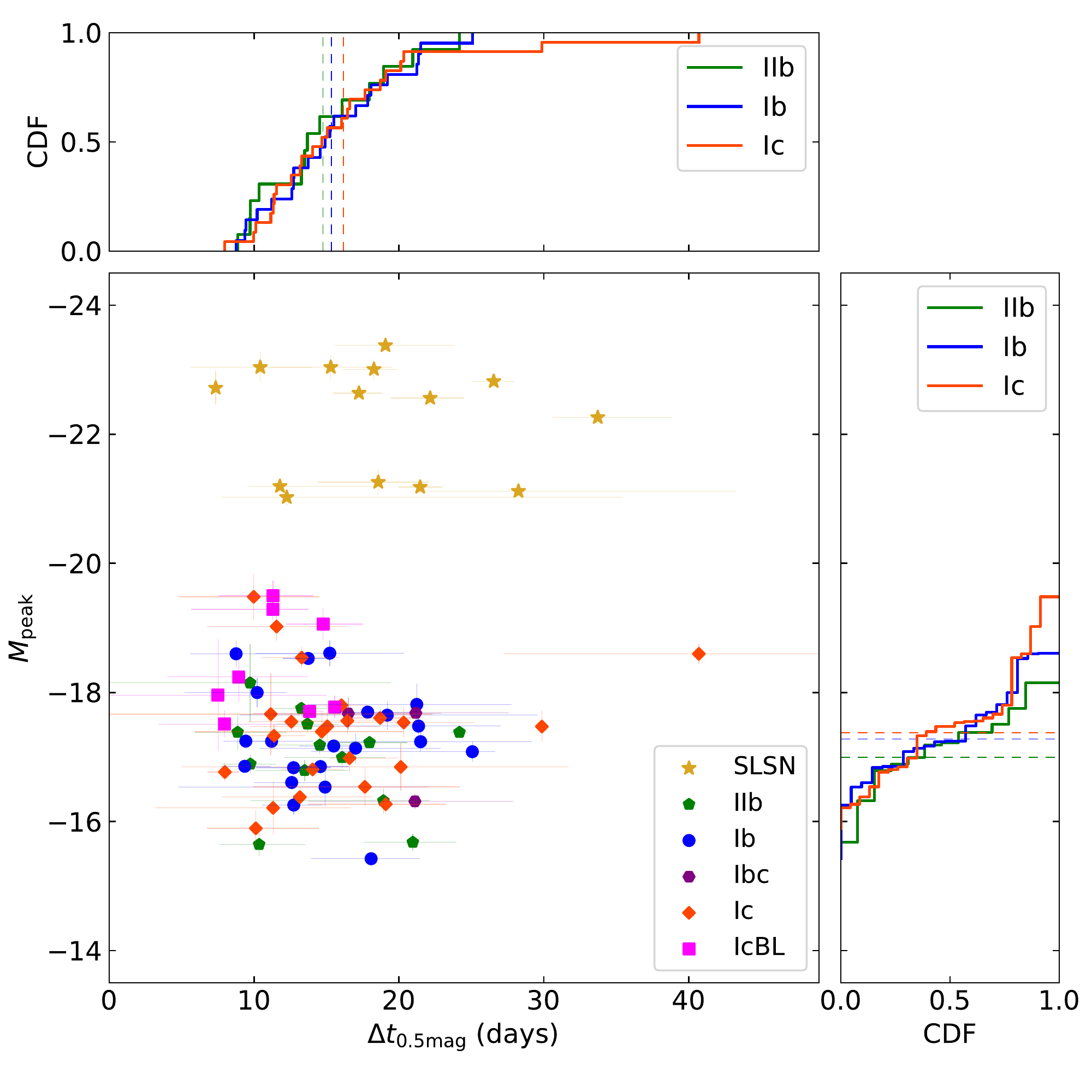}
\caption{A relation between the decline timescales and the peak luminosity.
The different colors show the different SN types of the SNe as shown in the legend.
The dashed lines in the CDFs show the average.}
\label{fig:deltat_Mpeak}
\end{figure*}

Figure \ref{fig:bol} shows the bolometric light curves of the 82 SNe
derived with the methods described in Section \ref{sec:data}.
The light curves have a variety of peak luminosities and time evolutions.
Some sharp features in the bolometric light curves are
due to the variation in the number of data points used in the SED fitting:
when the number of bands changes,
the best fit blackbody parameters can significantly vary, causing the sharp features.
From the bolometric light curves,
we estimate the decline timescales \deltat ~ and the peak bolometric magnitudes \Mpeak ~
(Table \ref{tab:summary}).
Figure \ref{fig:deltat_Mpeak} represents a relation between \deltat ~and \Mpeak.

The distribution of the decline timescales is concentrated from $\sim$ 10 days to $\sim$ 20 days.
On the other hand,
the distribution of the peak magnitudes is broad, from $\sim -14$ mag to $\sim -24$ mag.
All the SNe classified as SLSNe
in the database are more luminous than $-21$ mag
as expected from the definition.

It is noted that the distribution of the absolute magnitudes does not reflect the true distribution of SNe
since more luminous SNe have a bias to be observed more easily.
Also, although there seems a gap of the distribution, that is, less objects are located in around $-$ 20 mag,
it can also be considered as a bias that extremely luminous objects with the peak magnitudes $\lsim -21$ tend to get more attention
for follow-up observations.
In fact, previous systematic studies show that there is no gap in the luminosity distribution between normal SNe and SLSNe
\citep{Arcavi2016, DeCia2018, Perley2020, Prentice2021}.

Also, there is no clear correlation between the decline timescales and the peak magnitudes
showing a correlation coefficient of 0.07 without the SLSNe.
Furthermore, their distributions are not clearly different among Type IIb, Ib, and Ic SNe,
as shown in the cumulative distributions of \deltat ~ and \Mpeak ~ of Figure \ref{fig:deltat_Mpeak}.
P-values of \deltat ~ and \Mpeak ~ from K-S tests
for Type IIb vs. Ib SNe, Type Ib vs. Ic SNe, and Type Ic vs. IIb SNe are summarized in Table \ref{table:p}.
This indicates that their distributions are not distinguishable.

\subsection{\Mej ~ vs. \MNi}
\label{res:mass}

\begin{figure}[ht]
\begin{center}
  \begin{tabular}{c}
       \includegraphics[width= 0.97\linewidth]{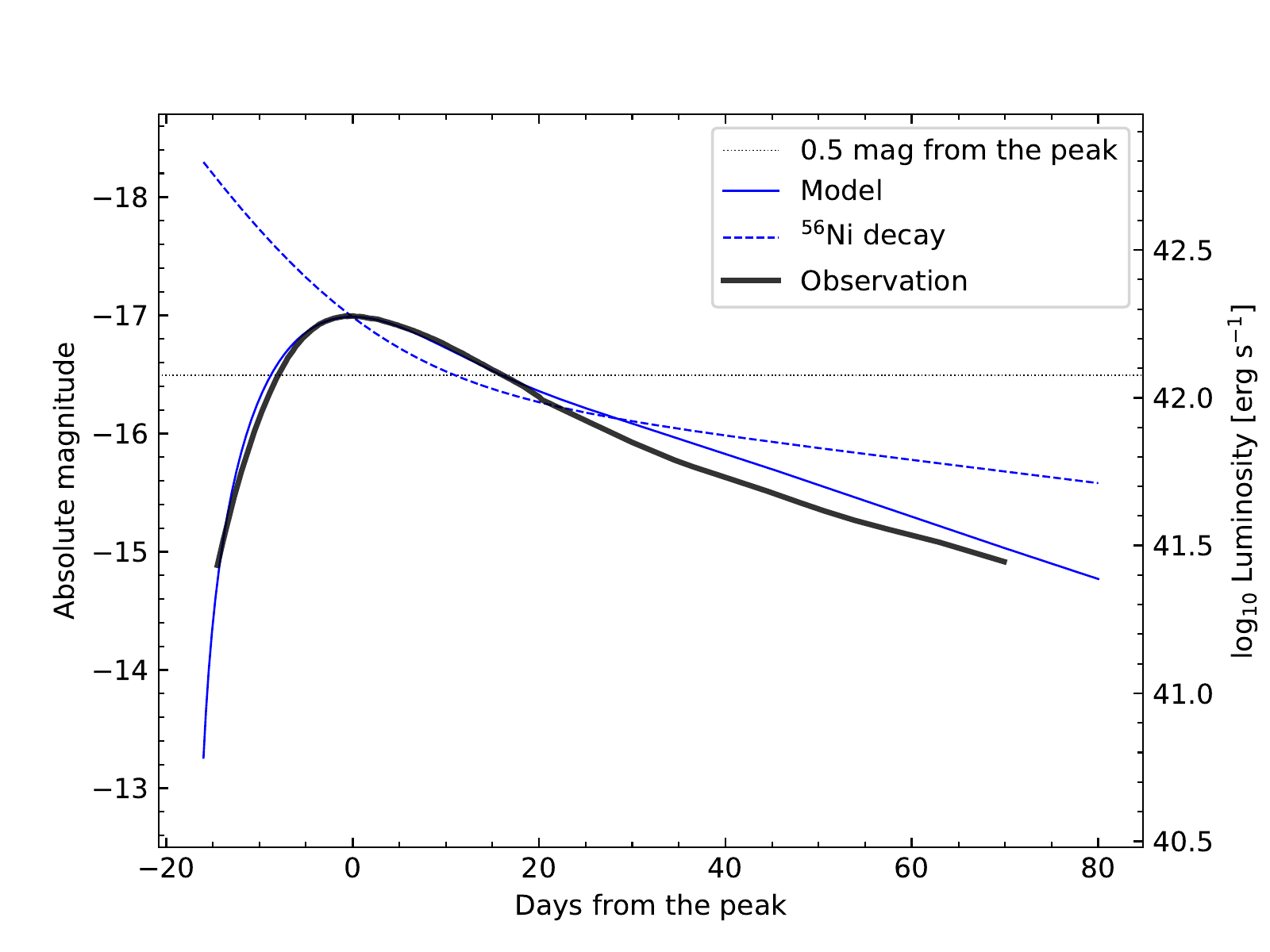}
  \end{tabular}  
\caption{An example of the comparison between the bolometric light curve constructed by observations (the black solid line, Type IIb SN 2006T)
and the corresponding semi-analytic model (the blue solid line)
with the deposition by the radioactive decay of \Nifs ~ and \Cofs ~ (the blue dashed line).
The luminosity of the observation and the semi-analytic model coincide
at their peaks and at the time when the magnitudes decline by 0.5 mag from the peaks (the black dotted line).
}
  \label{fig:LC_cmpr}
  \end{center}
\end{figure}

We estimate the ejecta mass \Mej ~ and the \Nifs ~ mass
from the decline timescale \deltat ~ and the peak magnitude \Mpeak, respectively,
using the semi-analytic model \citep{Arnett1982}.
Although we can also use, for example, $\Delta t_{\rm 1mag}$, which is the time for the light curve to decline by 1 mag from the peak,
to estimate the ejecta mass,
we have adopted \deltat ~ to obtain the timescale mainly determined by the photon diffusion.
For some of the light curves,  $\Delta t_{\rm 1mag}$ is affected by the \Cofs ~ tail.

First,  we estimate the ejecta mass \Mej ~ from the observed decline timescale \deltat.
For this purpose, we calculate a set of bolometric light curves using the semi-analytic model \citep{Arnett1982}
based on the prescription by \citet{Chatzopoulos2012}.
To calculate the light curves, we need the ejecta velocity \vej,
optical opacity $\kappa$ and gamma-ray opacity $\kappa_\gamma$,
and the ejecta mass \Mej.
In particular, \deltat ~ is sensitive to the ejecta velocity, the opacity in the ejecta, and the ejecta mass
as it reflects the diffusion timescale.
For the ejecta velocity, we adopt a constant, typical velocity $v_{\rm ej} = 10,000 {\rm ~km ~s^{-1}}$
($\pm~ 2,000 {\rm ~km ~s^{-1}}$) \citep[e.g.,][]{Lyman2016}
as observations show relatively homogenous velocity
(in Appendix A, we discuss the case when we fix the kinetic energy to $E_{\rm k} = 10^{51}$ erg).
It is noted that the variation of the velocity is not included in the error propagation.
Even though the error causes $\sim 20 ~ \%$ uncertainty of the ejecta mass,
this does not affect our conclusions (see Section \ref{dis2}).
For the optical opacity,
we use a constant opacity $\kappa= 0.07 {\rm ~ cm^2 ~g^{-1}}$ following previous studies \citep[e.g.,][]{Cano2013, Taddia2018},
which can reproduce the light curves of SNe well when the dominant opacity source is electron scattering \citep[e.g., ][]{Chevalier1992}.
For the gamma-ray opacity, we assume $\kappa_\gamma = 0.03 {\rm ~ cm^2 ~g^{-1}}$ \citep{Sutherland1984}.
This choice does not affect our results because we mainly use the light curve around the peak.
With these fixed parameters, we calculate bolometric light curves by varying the ejecta mass.
Then, by matching the calculated \deltat ~ and observed \deltat,
we derive the ejecta mass for each SN.
An example of the light curve comparison is shown in Figure \ref{fig:LC_cmpr}.

Next, we estimate the \Nifs ~ mass \MNi ~ from the peak magnitude (or the peak luminosity).
The \Nifs ~ mass is estimated from the peak luminosity of the SN
because the peak luminosity equals the instantaneous luminosity
generated by the radioactive decay of \Nifs ~ and \Cofs ~ \citep{Arnett1979}
as shown in Figure \ref{fig:LC_cmpr}.
The radioactive decay luminosity is expressed as follows \citep{Nadyozhin1994}:
\begin{eqnarray}
\label{eq:lum}
   L(t) = (L_{\rm Ni}   e^{-t/\tau_{\rm Ni}}
          + L_{\rm Co}  e^{-t/\tau_{\rm Co}} )          
           \left( \frac{M_{\rm Ni}}{M_\odot} \right),
\end{eqnarray}
where the $L_{\rm Ni}  = 6.45 \times 10^{43} {~ \rm erg ~ s^{-1}}$,
          $L_{\rm Co} = 1.45 \times 10^{43} {~ \rm erg ~ s^{-1}}$,
          $\tau_{\rm Ni} = 8.8$ days, and
          $\tau_{\rm Co} = 111.3$ days.
Substituting the peak time $t_{\rm peak}$ and the peak luminosity $L(t_{\rm peak})$ into Equation (\ref{eq:lum})
provides the \Nifs ~ mass.
The peak time $t_{\rm peak}$ is estimated from the decline timescale \deltat ~
with an assumption that the relation between the decline timescale and the peak time
follows that of the semi-analytic model \citep{Arnett1982, Chatzopoulos2012}.

\begin{figure*}[ht]
\centering
\includegraphics[width= 0.7\linewidth]{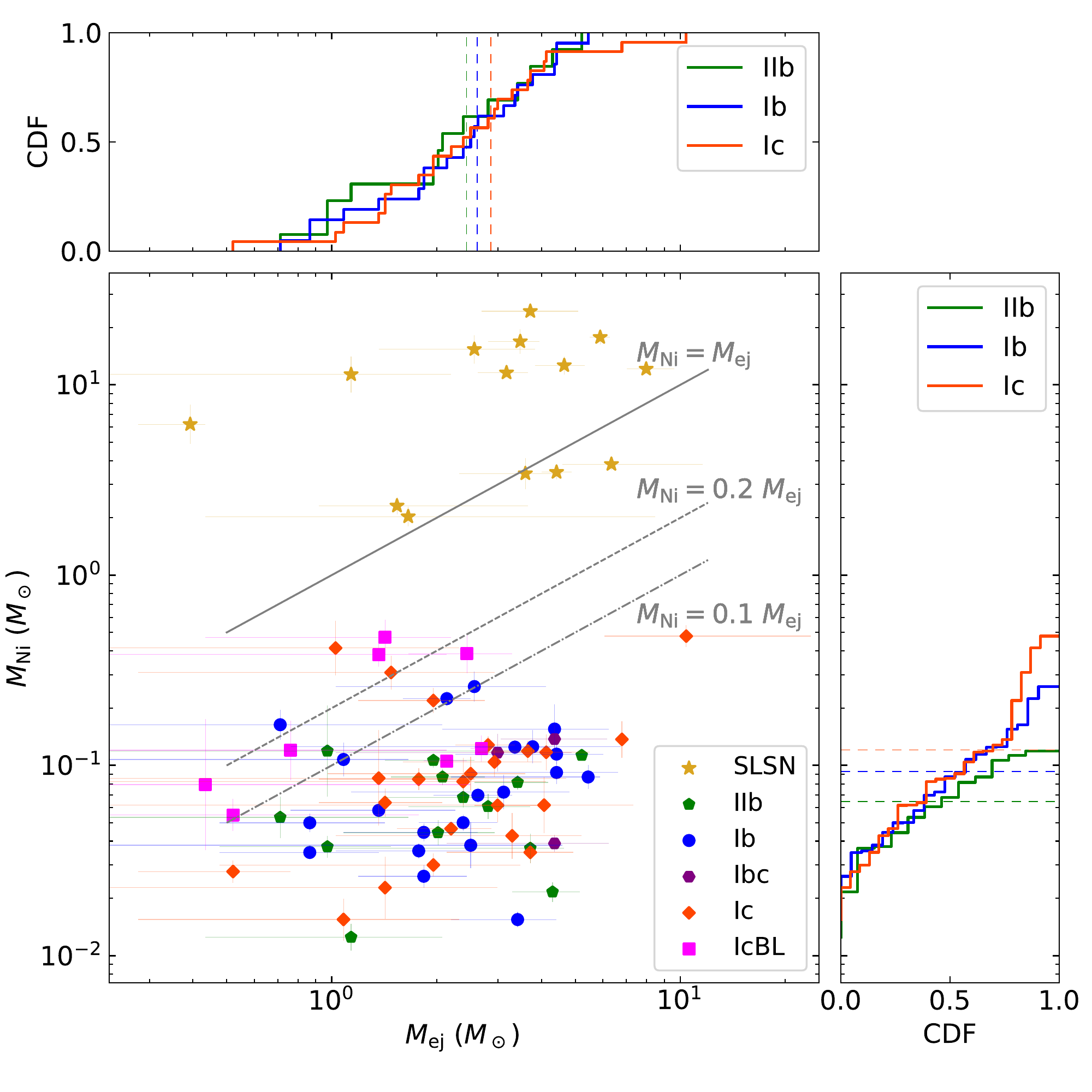}
\caption{
A comparison between ejecta masses and synthesized \Nifs ~ masses.
The different colors show different SN types as shown in the legend.
The gray lines represent
$M_{\rm Ni} = M_{\rm ej}, M_{\rm Ni} = 0.2 ~ M_{\rm ej} $, and $ M_{\rm Ni} = 0.1~ M_{\rm ej}$
from top to bottom.
The dashed lines in the CDF show the average.
}
\label{fig:ej_ni}
\end{figure*}

Figure \ref{fig:ej_ni} shows a relation between the ejecta masses \Mej ~ and the synthesized \Nifs ~ masses \MNi ~
for 82 SNe estimated in the methods above (Table \ref{tab:summary}).
Here, the derived \Nifs ~ masses have been divided by a correction factor of 1.4
since \citet{Dessart2016} has pointed out that the semi-analytic model \citep{Arnett1982} overestimates the \Nifs ~
by a factor of $\sim$ 1.4.
It is noted that \citet{Khatami2019} and \citet{Afsariardchi2021} have suggested a factor of $\sim$ 2 for the correction.
Because the \Nifs ~ mass estimated from the method in \citet{Khatami2019} is consistent
with that estimated from the tail of light curves,
the \Nifs ~ mass  can be considered as the lower limit of the \Nifs ~mass.
Thus, we divided the \Nifs ~ mass estimated from the semi-analytic model \citep{Arnett1982} by a factor of 1.4.

The distribution of the ejecta masses is concentrated
from $\sim 1$ \Msun ~ to $\sim 4$ \Msun ~ as shown in Figure \ref{fig:ej_ni}.
This result is consistent with previous studies \citep[e.g., ][]{Drout2011, Cano2013, Barbarino2021},
supporting our assumptions for the estimate of the ejecta mass.
From the distribution of the ejecta masses \Mej, pre-SN masses $M_{\rm preSN}$ are estimated to be $\sim 2.5$ \Msun ~ $ - \sim 5.5$ \Msun ~
with $M_{\rm preSN} = M_{\rm ej} + M_{\rm NS}$,
where $M_{\rm NS}$ the neutron star (NS) mass remaining after the SN explosion and the a typical NS mass is $\sim 1.4$ \Msun ~ \citep{Lattimer2012}.
These relatively small masses of pre-SNe suggest that
progenitor stars of almost all the SESNe originate from binary systems, 
as shown by previous studies \citep[e.g.,][]{Eldridge2008, Yoon2010, Sana2012, Sravan2019}.
The distributions of \Mej ~ and \MNi ~ do not depend on the SN types.
P-values from K-S tests for Type IIb vs. Ib SNe, Type Ib vs. Ic SNe, and Type Ic vs. IIb SNe
are summarized in Table \ref{table:p2}.

The \Nifs ~ masses of almost all the SLSNe exceed the ejecta masses
with the assumption that the heating source is the radioactive decay of \Nifs ~ even though it may not be the case.
The large \Nifs ~ masses suggest that the SLSNe require energy sources other than the radioactive decay of \Nifs.
Furthermore, although the \Nifs ~ masses of many normal SNe are less than 10 \% of the ejecta masses,
some normal SNe show high \Nifs ~ masses, $\sim 20 ~ \%$ of the ejecta masses.
In Section \ref{sec:dis},
we discuss necessary conditions to reproduce the \Nifs ~ mass derived for normal SNe.

\begin{table*}[t]
\begin{tabular}{cc}
  \begin{minipage}{0.40\hsize}
   \begin{center}
  \caption{P-values of the K-S test.}
  \label{table:p}
  \centering
  \begin{tabular}{c|ccc}
    \hline \hline
    Types & IIb \& Ib & Ib \& Ic & Ic \& IIb \\
    \hline
    $ \Delta t_{\rm 0.5mag}$ & 0.91 & 1.00 & 0.90 \\
    $ M_{\rm peak} $ & 0.93 & 0.56 & 0.25 \\
    \hline
  \end{tabular}
   \end{center}
  \end{minipage}
  \begin{minipage}{0.40\hsize}
  \begin{center}
   \caption{P-values of the K-S test.}
  \label{table:p2}
  \begin{tabular}{c|ccc}
    \hline \hline
    Types & IIb \& Ib & Ib \& Ic & Ic \& IIb \\
    \hline
    \Mej & 0.98 & 1.00 & 0.90 \\
    \MNi & 0.45 & 0.99 & 0.51 \\
    \hline
  \end{tabular}
  \end{center}
  \end{minipage}
\end{tabular}
\end{table*}

\section{Discussions}
\label{sec:dis}

\subsection{Hydrodynamics and nucleosynthesis calculations}
\label{dis1}

In this section,
we investigate what conditions of SN explosion can reproduce the relation
between the ejecta mass and the synthesized \Nifs ~ mass
by performing hydrodynamics and nucleosynthesis calculations.
Since the synthesized \Nifs ~ mass depends not only on the explosion timescales but also on structures of progenitor stars,
we take them into account to discuss the explosion timescales and synthesized \Nifs ~ masses.
Finally, we compare results of the calculations with the observations.

\subsubsection{Method of calculations}
\label{dis:method}

We employ a one-dimensional Lagrangian hydrodynamics calculation code, blcode \citep{Morozova2015}
and 21-isotopes nucleosynthesis \footnote{\url{http://cococubed.asu.edu/code_pages/burn_helium.shtml}} in our calculations,
following the modeling of the explosion performed in \citet{Sawada2019}.
It is noted that production of \Nifs ~ is slightly overestimated in the case of 21-isotopes \citep{Timmes1999}.
However, this does not affect the final conclusions
because we discuss the lower limit of the explosion timescale, that is, the upper limit of the \Nifs ~ mass.
Thus, it is more conservative to consider the case that can synthesize more \Nifs ~ mass.

Basic equations of hydrodynamics to be solved are following:
\begin{eqnarray}
    \pd{r}{m} &=& \frac{1}{4\pi r^2 \rho}
    \\
    \ld{v}{t} &=& -\frac{Gm}{r^2} 
                - 4\pi r^2 \pd{P}{m}
    \\
    \ld{\epsilon}{t} &=& -P \ld{}{t} 
                       \left(
                       \frac{1}{\rho}
                       \right)
                       + \dot \epsilon_{\rm in}
\end{eqnarray}
where $\dot \epsilon_{\rm in}$ is the energy injection rate per mass.
Here, the term of the energy generation by the nucleosynthesis
is ignored because the energy is negligible
compared with the energy injection $\epsilon_{\rm in}$ in this work.
The energy injection rate $\dot \epsilon_{\rm in}$ is described as
\begin{eqnarray}
\label{eq:Ein}
    \dot \epsilon_{\rm in}
    = \frac{E_{\rm k} - (E_{\rm bind} + E_{\rm int})  }{t_{\rm grow}}     \frac{1}{\Delta M}
\end{eqnarray}
so that the SN ejecta finally expand with the kinetic energy $E_{\rm k}$.
Here, $E_{\rm bind}$ is a binding energy of a progenitor star,
$E_{\rm int}$ is an internal energy of the progenitor,
$\Delta M$ is a mass in each mesh
($\Delta M = (M_{\rm core} - 1.4 ~ M_{\odot})/N_{\rm grid}$; see Section 4.1.2 for $M_{\rm core}, N_{\rm grid} = 1000$),
and $t_{\rm grow}$ is a timescale of an explosion.
The energy is injected just outside of the inner boundary.
It is noted that all the models are assumed to always explode with the input energy
even though some progenitors may directly collapse to black holes without explosion
\citep[e.g.,][]{MacFadyen1999, Ugliano2012, Ertl2020}.

\subsubsection{Parameters of calculations}
\label{dis:progenitor}

We parameterize the ejecta mass \Mej, which is related to the mass of the progenitor star,
and the explosion timescale \tgrow, with which the ejecta reach the final kinetic energy $E_{\rm k}$
as shown in Equation \ref{eq:Ein}.
This is because the synthesized \Nifs ~ mass depends
both on the structure of the progenitor star and the explosion timescale.
For the determination of the progenitor star from the ejecta mass, we adopt stellar evolution models by \citet{Sukhbold2016}.
Note that these are single star models
although most of SESNe are considered to originate from binary systems
\citep[e.g.,][]{Maund2004, Eldridge2008, Yoon2010, Sana2012, Sravan2019}.
The use of single star models is justified
because the inner structures of the progenitors from single stars are similar to those from binary systems
when their core masses are similar \citep[e.g.,][]{Sukhbold2016, Vartanyan2021}.

For the ejecta mass, $M_{\rm ej} = 1 ~ M_\odot - 10 ~ M_\odot$ 
are used to cover the range of the ejecta masses estimated from the observations, with a step of $1 ~ M_\odot$.
The ejecta mass plus the NS mass provides the corresponding pre-SN mass $M_{\rm preSN}$ (namely the core mass $M_{\rm core}$)
from this equation ($M_{\rm preSN} = M_{\rm core} = M_{\rm ej} + M_{\rm NS}$).
Here, a typical mass of the NS mass, 1.4 \Msun ~ is adopted \citep{Lattimer2012}.
The uncertainty of the \Nifs ~ mass due to the NS mass is discussed in Appendix B,
but our conclusions are not affected by the NS mass.
Since the core mass is discretized following a model with the mass resolution of the ZAMS masses (1 \Msun),
the model with the nearest core mass is selected.
Here, two progenitor stars, the progenitors with He layer and without He layer,
are used to consider Type Ib and Ic SNe, respectively.
The parameters of the progenitors used in our calculations are summarized in Table \ref{table:data_type}.

For the explosion timescale, $t_{\rm grow} = [0.01, 0.03, 0.1, 0.3, 1.0] ~ {\rm sec}$ are tested.
The lower bound approximately corresponds to a free fall time of a iron core,
which represents a prompt explosion.
The upper bound is set to be 1 sec because
almost no \Nifs ~ is synthesized with $t_{\rm grow} > 1$ sec.

With the explosion timescale, the ejecta reach the final kinetic energy $E_{\rm k}$
expressed as $E_{\rm k} = (3/10)M_{\rm ej} {v_{\rm ej}}^2$.
This expression assumes homologously expanding ejecta ($R = v_{\rm ej}t$)
with a uniform density ($\rho = 3 M_{\rm ej} /4\pi R^3$)
even though the actual density has a structure.
We here fixed a typical velocity ($v_{\rm ej} = 10,000 {\rm ~km ~ s^{-1}}$)
as in the parameter estimation in Section \ref{res:mass}.
In Appendix A, we show the results fixing the kinetic energy to $E_{\rm k}= 10^{51}$ erg.

\begin{table*}[t]
  \caption{A summary of progenitor star models}
  \label{table:data_type}
  \centering
  \begin{tabular}{c|cccc}
    \hline \hline
    Model & $M_{\rm ZAMS} \ (M_\odot)$  & $E_{\rm bind} \ (10^{51} \ {\rm erg})$ 
    & $E_{\rm int} \ (10^{51} \ {\rm erg})$  & $E_{\rm k} \ (10^{51} \ {\rm erg})$
    \\
    \hline
He 1  &  10  &  -5.25  &  4.62  &  0.6  \\ 
He 2  &  13  &  -5.32  &  4.58  &  1.2  \\ 
He 3  &  15  &  -5.72  &  5.17  &  1.8  \\ 
He 4  &  18  &  -5.88  &  5.29  &  2.4  \\ 
He 5  &  20  &  -6.26  &  5.58  &  3.0  \\ 
He 6  &  23  &  -7.48  &  6.66  &  3.6  \\ 
He 7  &  25  &  -7.74  &  6.89  &  4.2  \\ 
He 8  &  28  &  -8.23  &  7.35  &  4.8  \\ 
He 9  &  30  &  -7.62  &  6.76  &  5.4  \\ 
He 10  &  32  &  -8.14  &  7.19  &  6.0  \\ 
\hline 
CO 1  &  13  &  -5.40  &  3.91  &  0.6  \\ 
CO 2  &  16  &  -5.63  &  4.02  &  1.2  \\ 
CO 3  &  19  &  -5.95  &  3.86  &  1.8  \\ 
CO 4  &  21  &  -5.96  &  3.96  &  2.4  \\ 
CO 5  &  24  &  -8.04  &  3.67  &  3.0  \\ 
CO 6  &  26  &  -7.49  &  3.70  &  3.6  \\ 
CO 7  &  28  &  -8.51  &  4.00  &  4.2  \\ 
CO 8  &  31  &  -8.02  &  3.76  &  4.8  \\ 
CO 9  &  33  &  -8.74  &  3.65  &  5.4  \\ 
CO 10  &  35  &  -9.63  &  3.62  &  6.0  \\ 
\hline 
  \end{tabular}
\end{table*}

\subsubsection{Synthesized \Nifs ~ mass}
\label{dis1:Ni}

\begin{figure*}[ht]
\begin{center}
  \begin{tabular}{c}
    \begin{minipage}{0.5\hsize}
      \begin{center}
       \includegraphics[width= 0.65 \linewidth, bb=120 0 500 560]{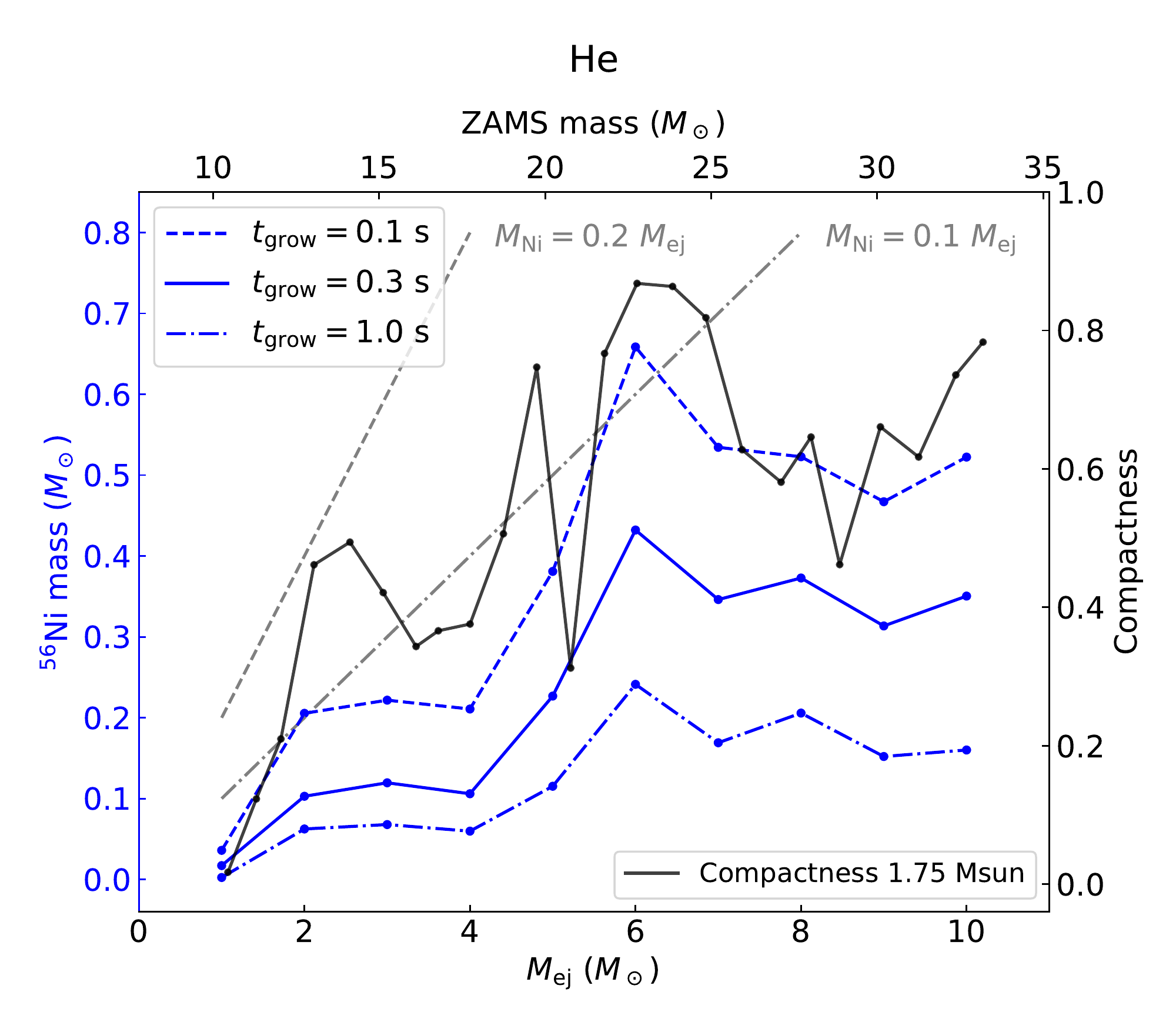}
        \end{center}
    \end{minipage}    
     \begin{minipage}{0.5\hsize}
      \begin{center}
       \includegraphics[width= 0.65 \linewidth, bb=120 0 500 560]{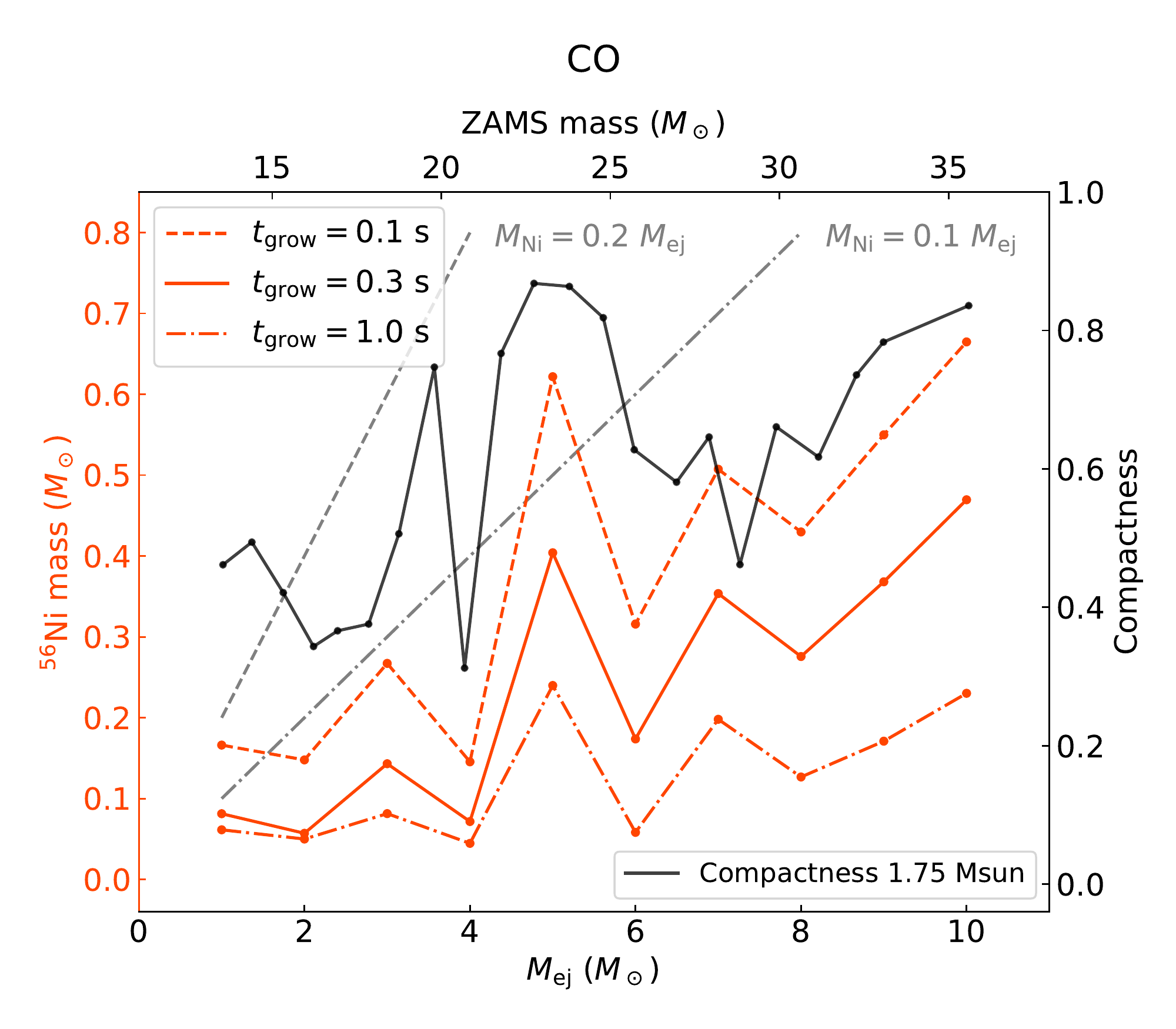}
       \end{center}
    \end{minipage}
  \end{tabular}  
\caption{A relation between the ejecta mass and the synthesized \Nifs ~ mass calculated with the various explosion timescales.
Left panel: results from He cores. Right panel: results from CO cores.
The different types of the colored lines show the different explosion timescales as shown in the legend.
The black lines show compactness at 1.75 \Msun ~ in mass coordinate.
The dashed-dotted and dashed gray lines are the lines showing
$M_{\rm Ni} = 0.1 \ M_{\rm ej}$ and $M_{\rm ej} = 0.2 \ M_{\rm ej}$, respectively.}
  \label{fig:ni}
  \end{center}
\end{figure*}

Table \ref{table:MNi} summarizes the synthesized \Nifs ~ mass
for each progenitor model with the five explosion timescales.
Figure \ref{fig:ni} represents the ejecta mass and the \Nifs ~ mass with the explosion timescale of 0.1, 0.3, and 1.0 sec, 
showing a comparison with the compactness parameter
\citep[$\xi_{\rm M = 1.75M_\odot} = (M/M_\odot)/(R(M)/1000 {\rm ~ km})$,][]{OConnor2011}.
We do not show the results of explosion timescale of 0.01 and 0.03 sec
because the synthesized \Nifs ~ masses of these timescales are the almost same as those of 0.1 sec as shown in Table \ref{table:MNi}.
Here, it should be noted that
Figure \ref{fig:ni} shows the amount of \Nifs ~ mass newly synthesized by explosive nucleosynthesis,
which corresponds to the maximum ejectable amount of \Nifs.
\begin{table*}[t]
  \caption{Summary of the synthesized \Nifs ~ masses (in \Msun) for each progenitor model and explosion timescale, $t_{\rm grow}$.}
  \label{table:MNi}
  \centering
  \begin{tabular}{l|ccccc}
    \hline \hline
          &&& {$t_{\rm grow}$ (s)} && \\
Model & 0.01 & 0.03 & 0.1 & 0.3 & 1.0 \\
\hline
He 1 & 0.048 & 0.044 & 0.036 & 0.017 & 0.002 \\ 
He 2 & 0.237 & 0.235 & 0.205 & 0.103 & 0.062 \\ 
He 3 & 0.254 & 0.249 & 0.222 & 0.120 & 0.068 \\ 
He 4 & 0.232 & 0.229 & 0.211 & 0.106 & 0.060 \\ 
He 5 & 0.404 & 0.400 & 0.381 & 0.227 & 0.115 \\ 
He 6 & 0.689 & 0.677 & 0.659 & 0.432 & 0.241 \\ 
He 7 & 0.555 & 0.548 & 0.534 & 0.346 & 0.169 \\ 
He 8 & 0.530 & 0.533 & 0.523 & 0.373 & 0.206 \\ 
He 9 & 0.456 & 0.461 & 0.467 & 0.314 & 0.152 \\ 
He 10 & 0.508 & 0.512 & 0.522 & 0.350 & 0.160 \\ 
\hline 
CO 1 & 0.215 & 0.210 & 0.166 & 0.081 & 0.061 \\ 
CO 2 & 0.179 & 0.175 & 0.148 & 0.057 & 0.050 \\ 
CO 3 & 0.285 & 0.281 & 0.267 & 0.143 & 0.081 \\ 
CO 4 & 0.166 & 0.163 & 0.146 & 0.072 & 0.045 \\ 
CO 5 & 0.656 & 0.642 & 0.622 & 0.404 & 0.240 \\ 
CO 6 & 0.324 & 0.326 & 0.316 & 0.174 & 0.058 \\ 
CO 7 & 0.518 & 0.519 & 0.507 & 0.354 & 0.198 \\ 
CO 8 & 0.422 & 0.430 & 0.430 & 0.276 & 0.127 \\ 
CO 9 & 0.547 & 0.545 & 0.550 & 0.368 & 0.171 \\ 
CO 10 & 0.672 & 0.674 & 0.665 & 0.470 & 0.230 \\ 
\hline    
  \end{tabular}
\end{table*}

In Table \ref{table:MNi} and Figure \ref{fig:ni},
three trends of the \Nifs ~ mass are seen:
(1) the synthesized \Nifs ~ mass decreases with the increase of the explosion timescale,
(2) the \Nifs ~ mass correlates with the compactness, and
(3) the \Nifs ~ roughly increases with the increase of the ejecta mass.
These trends of the \Nifs ~ mass can be explained as follows.

\Nifs ~ is synthesized around the core of the progenitor stars at the temperature
$T \gtsim 5 \times 10^9$ K \citep[e.g.,][]{Woosley2002, Seitenzahl2009} after the shock propagation.
In the shocked region, the kinetic energy $E_{\rm k}$ is comparable to the internal energy $E_{\rm int}$.
Also, the region for \Nifs ~  production is hot enough to be radiation dominant.
Thus, the radius $R$ of the temperature at which \Nifs ~ is synthesized can be estimated as below:
\begin{gather}
    E_{\rm k}
    \simeq E_{\rm int} \simeq E_{\rm rad} = aT^4 \frac{4\pi}{3} R^3 \\
\label{eq:R}
    R 
    \sim
     4 \times 10^8 {\rm ~ cm} 
    \left(
    \frac{E_{\rm k}}{10^{51} {\rm ~ erg}}
    \right) ^{1/3}
    \left(
    \frac{T}{5\times 10^9 {\rm ~ K}}
    \right) ^{-4/3}
\end{gather}
where $E_{\rm rad}$ is the radiation energy.
Therefore, the synthesized \Nifs ~ mass is approximately determined by the mass enclosed in this radius.

The first trend that the decrease of the \Nifs ~ mass with the increase of the explosion timescale
also shown in previous studies \citep[e.g., ][]{Sawada2019} is interpreted as follows.
The radius in Equation \ref{eq:R} is estimated for the case of the prompt explosion.
In the case of a slow explosion, the SN starts to expand adiabatically due to the energy obtained initially.
This causes the decrease of the internal energy, which in turn reduces the radius above.
As a result, the amount of the synthesized \Nifs ~ decreases with the explosion timescale.
Explosion with the timescales less than 0.1 sec shows almost constant \Nifs ~ masses.
This is due to the timescale of the shockwave propagation.
It takes $\sim 0.4$ sec for the shockwave to propagate the radius
where the \Nifs ~ is synthesized ($R \ltsim 4 \times 10^8 {~\rm cm}$)
with the shock velocity of $v_{\rm sh} = 10,000 {\rm ~ km ~ s^{-1}}$.
Therefore, when the explosion timescale $t_{\rm grow}$ is considerably shorter than this timescale,
the explosion is considered to be prompt, producing the almost constant \Nifs ~ masses.

Second, the variation of the \Nifs ~ mass with the ejecta mass follows the compactness of the core.
In the cases of the higher compactness, the mass where \Nifs ~ is synthesized ($R \ltsim 4 \times 10^8 {\rm ~cm}$) increases.
This enhances the synthesized \Nifs ~ mass.

Third,
there is an overall trend that the \Nifs ~ mass roughly increases with the increase of the ejecta mass.
This is because the kinetic energy is assumed to be proportional to the ejecta mass in our calculations.
For the fixed kinetic energy ($E_{\rm k} = 10^{51}$ erg),
this overall trend of the \Nifs ~ mass with the increase of the ejecta mass is reduced (see Appendix A).

Our results give the highest luminosity for \Nifs-powered SNe
since the synthesized \Nifs ~ mass is the maximum ejectable amount of \Nifs.
No model except for the models of the CO core
with the ejecta mass of 1 \Msun ~ and with the explosion timescale of 0.01 and 0.03 sec synthesize $M_{\rm Ni} > 0.2 ~ M_{\rm ej}$.
Even though the inner boundary is changed to $M = 1.2 $ \Msun,
only the models of the CO core with the ejecta mass of 1 \Msun ~ and with the explosion timescale of $\leq$ 0.1 sec
can synthesize \MNi \gtsim 0.2 \Mej ~ (see Appendix B).
This indicates that the maximum luminosity that normal SNe powered by \Nifs ~ can reach
is expressed as $M_{\rm Ni} \ltsim 0.2 \  M_{\rm ej}$.
The objects with the \Nifs ~ mass exceeding this limit would suggest the existence of other power sources.

\subsection{Constraints to the explosion timescale}
\label{dis2}

\begin{figure*}[ht]
\begin{center}
  \begin{tabular}{c}
    \begin{minipage}{0.5\hsize}
      \begin{center}
        \includegraphics[width= 0.97\linewidth,  bb=7 7 460 580]{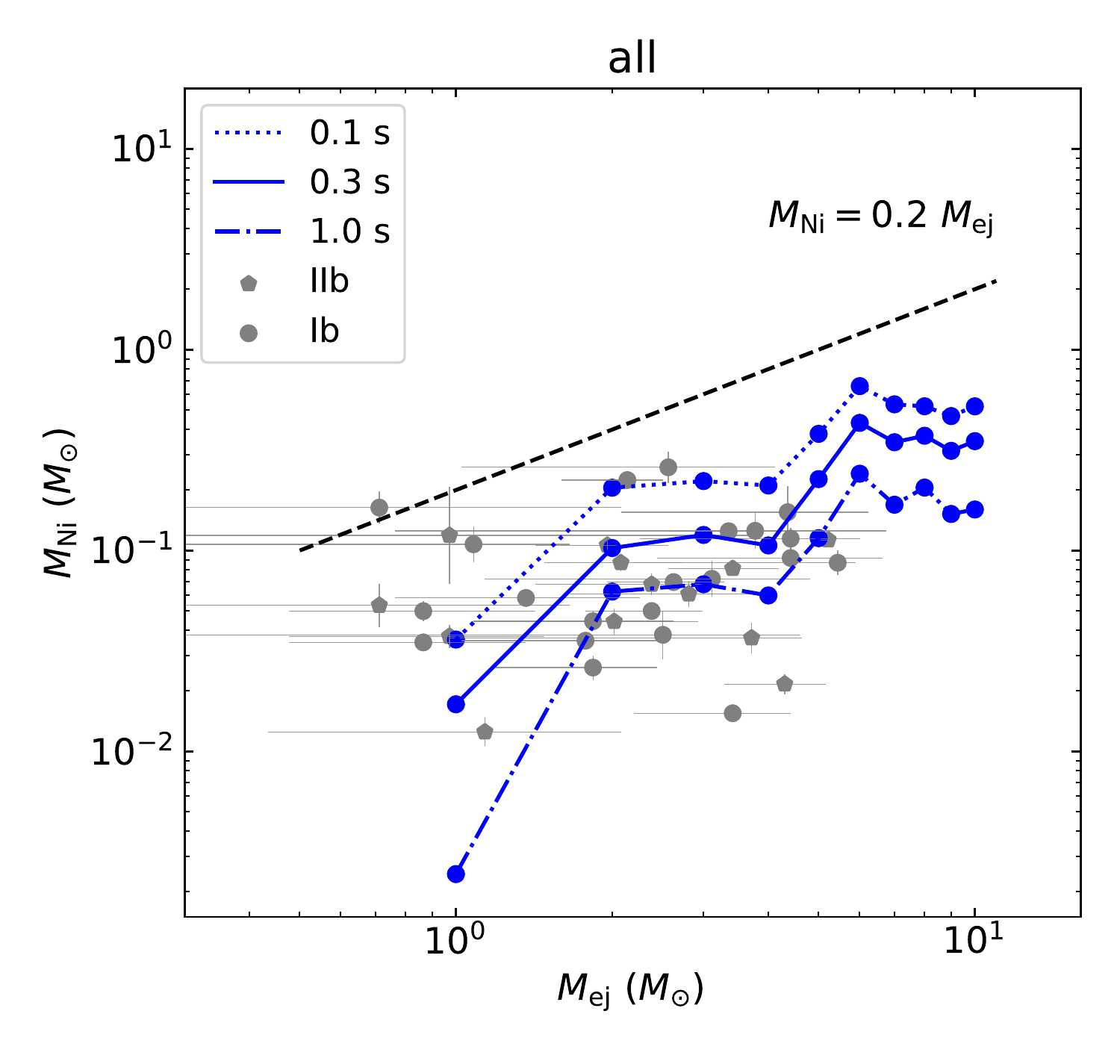}
        \end{center}
    \end{minipage}    
     \begin{minipage}{0.5\hsize}
      \begin{center}
       \includegraphics[width= 0.97\linewidth, bb=7 7 460 580]{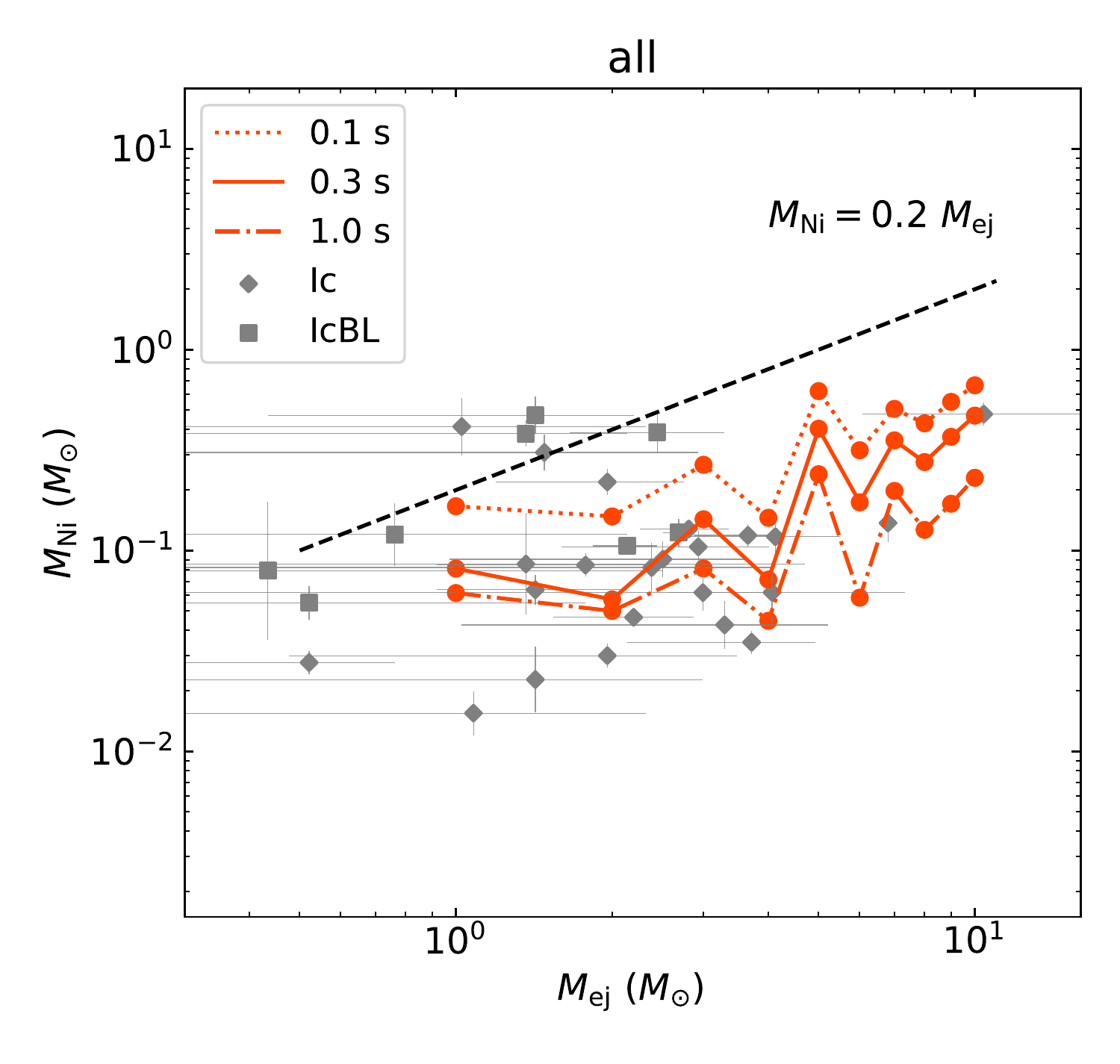}
       \end{center}
    \end{minipage}    
  \end{tabular}  
  \caption{A comparison of the ejecta masses and the synthesized \Nifs ~ masses.
The gray dots represent the observations,
and the colored solid, dashed and dashed-and-dotted lines show results of the calculations
with explosion timescales of 0.1 sec, 0.3 sec and 1.0 sec, respectively.
The black dashed line shows a condition where $M_{\rm Ni} = 0.2 \ M_{\rm ej}$.
Left panel: Type IIb and Type Ib SNe with the calculations using the progenitor models of the He cores.
Right panel: Type Ic and Type Ic BL SNe with the calculations using the progenitor models of the CO cores.
}
  \label{fig:cmpr}
  \end{center}
\end{figure*}

Finally, we compare our results of the calculations with the observations
to give constraints on the explosion timescale.
Figure \ref{fig:cmpr} shows a relation between the ejecta masses and the synthesized \Nifs ~ masses
estimated from both the observations and the calculations.
Since the \Nifs ~ masses are almost constant
when the explosion timescales are less than 0.1 sec (see Section \ref{dis1:Ni}),
we only show the results for $t_{\rm grow} = $ 0.1, 0.3, and 1.0 sec in Figure \ref{fig:cmpr}.
The observational results of Type IIb and Ib SNe (Type Ib-like SNe) and Type Ic and Ic-BL SNe (Type Ic-like SNe)
are compared to the results of the progenitor models of the He core and the CO core, respectively.
There are 34 Type Ib-like SNe and 31 Type Ic-like SNe in the diagrams.
The SLSNe are excluded from the sample since their luminosity cannot be reproduced by the radioactive decay of \Nifs.

In Figure \ref{fig:cmpr},
when the \Nifs ~ masses of the observed SNe are below the calculated \Nifs ~ masses,
the observed SNe can be explained by the calculations.
This is because the synthesized \Nifs ~ mass is the upper limit of the ejected \Nifs ~ mass with the SN ejecta
as already mentioned in Section \ref{dis1:Ni}.
As shown in Figure \ref{fig:cmpr}, 
the \Nifs ~ masses of almost all the observed SNe can be explained in the cases of $t_{\rm grow} = $ 0.1 sec.
On the other hand, with $t_{\rm grow} = $ 1.0 sec, 
\Nifs ~ masses of more than half of the SNe are not reproduced.
This result is not affected by the uncertainties of the ejecta masses
even though we include additional 20 \%  error caused by the uncertainty of the ejecta velocity (Section \ref{res:mass}).

Here, we discuss three systematic uncertainties in our analysis.
First, we consider the accuracy of the \Nifs ~ mass estimated from the semi-analytic model \citep{Arnett1982}.
In Arnett's rule, the distribution of the heating source is assumed be the same as that of the density and the temperature.
This is known to cause overestimate of the \Nifs ~ mass
especially in cases of longer timescales of light curves \citep[e.g.,][]{Khatami2019}.
As mentioned in Section \ref{res:mass},
the \Nifs ~ masses shown in Figure \ref{fig:ej_ni} are already divided by a factor of 1.4
to account for the effect.
This has relaxed the condition of the explosion timescale $t_{\rm grow}$,
i.e., the longer timescale is acceptable.

Second, we have ignored the extinction of host galaxies, as mentioned in Section \ref{data3}.
If the extinction is $A_V \sim 0.6$ for Type IIb and Ib SNe, $A_V \sim 1$ for Type Ic SNe \citep{Stritzinger2018},
or $A_V \sim 0.9$ for SESNe \citep{Zheng2022}, 
the actual synthesized \Nifs ~ masses are larger by a factor of $\sim 1.7$, $\sim 2.5$,  or $\sim 2.3$, respectively.
This requires even shorter explosion timescales $t_{\rm grow}$.
Therefore, ignoring extinction of host galaxies
results in an underestimation of \Nifs ~ mass,
and thus gives a conservative constraint to the explosion timescale $t_{\rm grow}$.

\begin{figure*}[ht]
\begin{center}
  \begin{tabular}{c}
    \begin{minipage}{0.5\hsize}
      \begin{center}
        \includegraphics[width= 0.97\linewidth, bb=7 7 460 580  ]{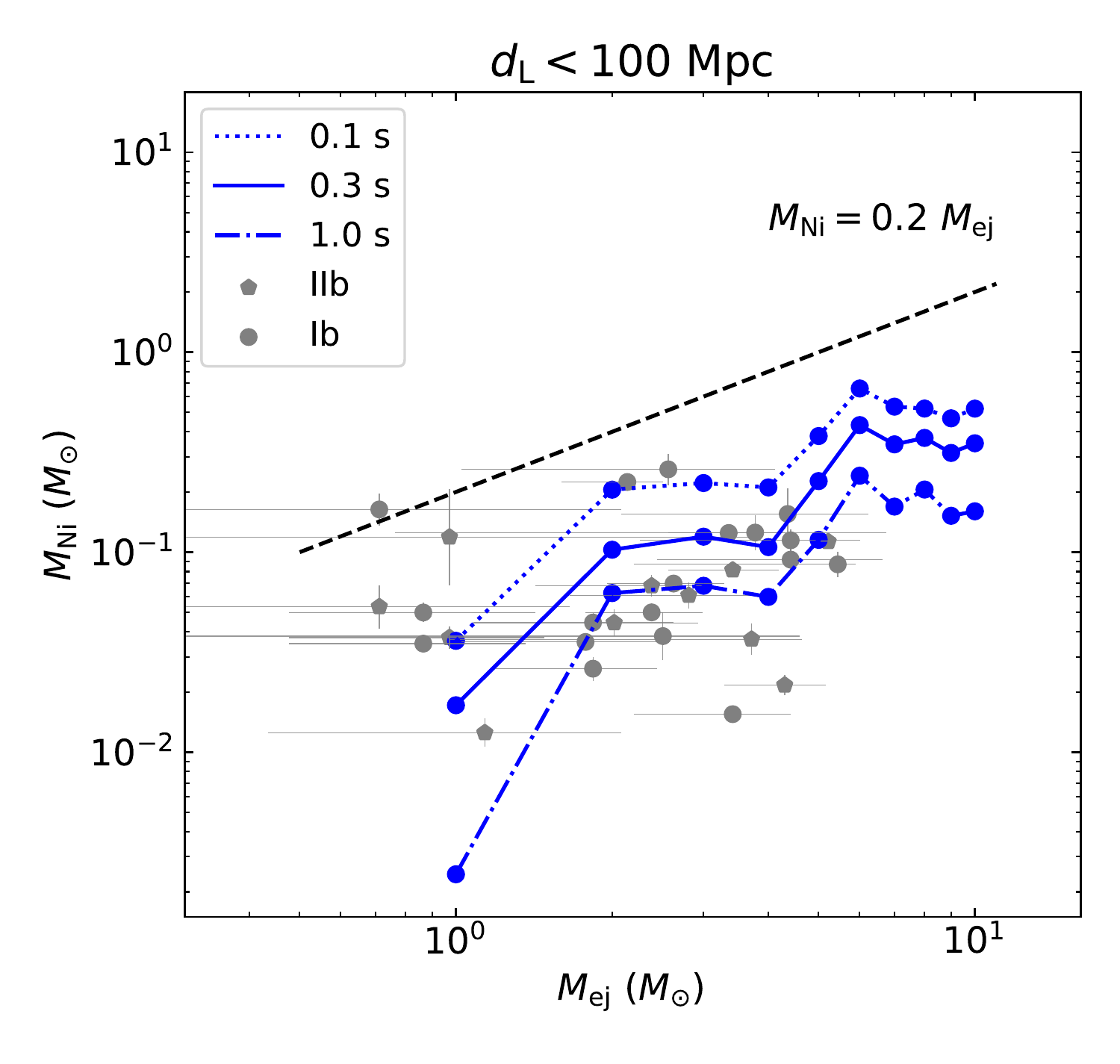}
        \end{center}
    \end{minipage}    
     \begin{minipage}{0.5\hsize}
      \begin{center}
       \includegraphics[width= 0.97\linewidth, bb=7 7 460 580 ]{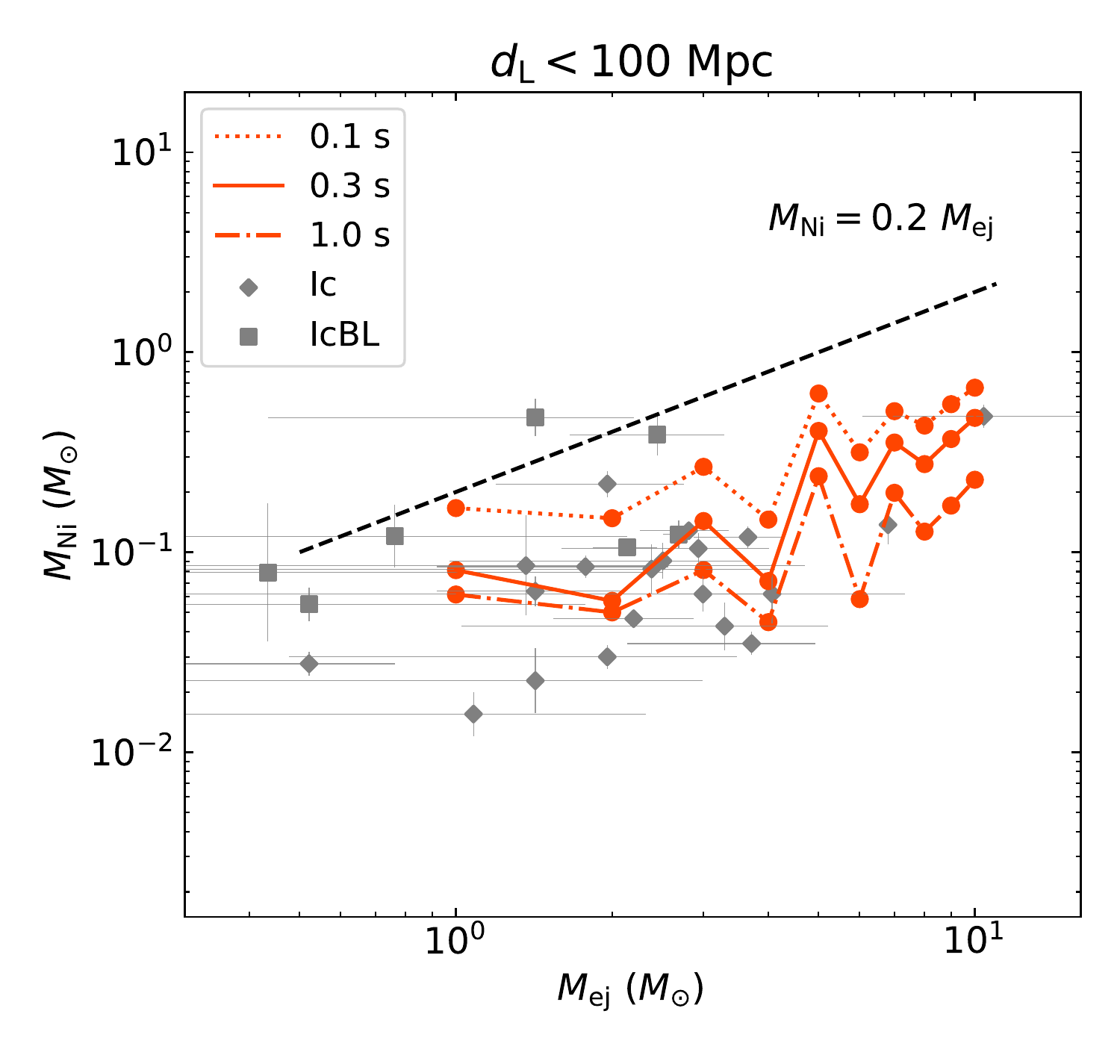}
       \end{center}
    \end{minipage}    
  \end{tabular}  
  \caption{Same as Figure \ref{fig:cmpr} but for the volume-limited sample within 100 Mpc.}
  \label{fig:cmpr_vlim}
  \end{center}
\end{figure*}

Third, we change the way to select the sample objects.
Since more luminous objects are detectable at larger distances,
we limit the distance of the samples to  remove the observational bias.
We adopt the luminosity distance of $d_{\rm L} =$ 100 Mpc as the threshold.
The numbers of Type Ib-like and Type Ic-like SNe within $d_{\rm L} =$ 100 Mpc decreased to 29 and 27, respectively.
Figure \ref{fig:cmpr_vlim} shows this relation.
Again, this gives conservative constraints to the explosion timescale $t_{\rm grow}$.

\begin{figure*}[ht]
\begin{center}
  \begin{tabular}{c}
    \begin{minipage}{0.5\hsize}
      \begin{center}
        \includegraphics[width= 0.97\linewidth, bb=7 7 460 580  ]{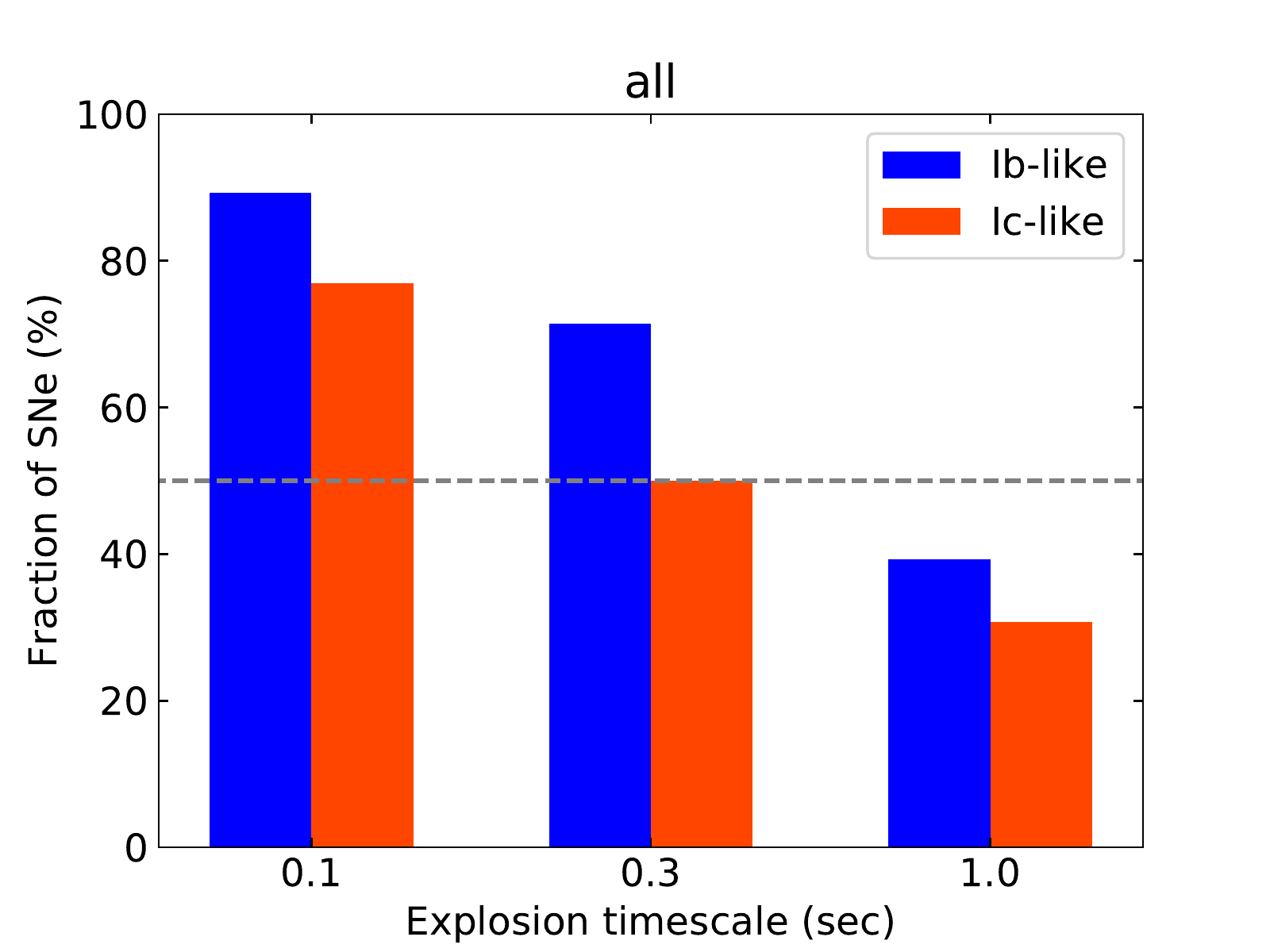}
        \end{center}
    \end{minipage}    
     \begin{minipage}{0.5\hsize}
      \begin{center}
       \includegraphics[width= 0.97\linewidth, bb=7 7 460 580  ]{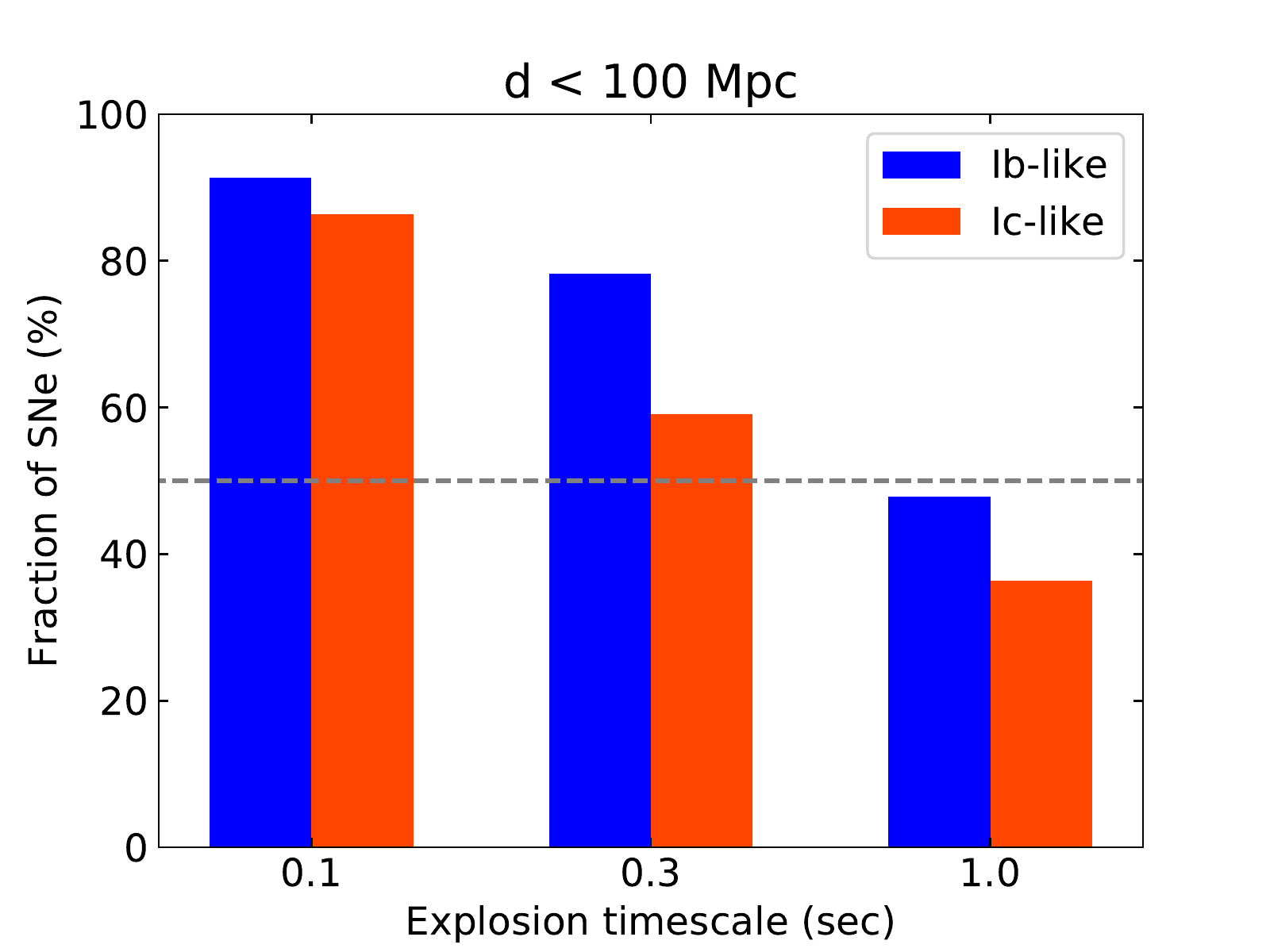}
       \end{center}
    \end{minipage}    
  \end{tabular}  
  \caption{Fractions of SNe whose \Nifs ~ mass can be explained with each explosion timescale.
  The blue and red colors show Type Ib-like (IIb and Ib) and Type Ic-like (Ic and Ic BL) SNe, respectively.
  Left: all the sample. Right: volume-limited sample within 100 Mpc.
  The horizontal gray dashed lines show 50 \% fraction.}
  \label{fig:percent}
  \end{center}
\end{figure*}

Figure \ref{fig:percent} summarizes fractions that the synthesized \Nifs ~ masses can be explained by each explosion timescale
for all the sample and for the volume-limited sample.
Here, we note that we excluded the SNe with the ejecta mass of $< 1$ \Msun ~ or  $> 10$ \Msun ~
because the progenitor models of our calculations do not cover the mass range.
The fraction that the \Nifs ~ mass can be explained by our calculations increases by $\sim 5 ~ \% $
from all the sample to the volume-limited sample
since the number of SNe with larger \Nifs ~ mass decreased.
The synthesized \Nifs ~ mass from our calculations of the explosions with the timescale of 1.0 sec (“slow” explosions preferred by ab-initio calculations)
cannot explain the \Nifs ~ mass of  $50 ~ \%$ of SNe in the volume-limited sample.

To explain \Nifs ~ mass of the majority of SNe even in the volume-limited sample,
the explosion timescale should be shorter than 0.3 sec.
As discussed above, we have divided the \Nifs ~ mass by 1.4 (Section \ref{res:mass})
and ignored the host extinction (Section \ref{data3}) to give conservative estimate of \Nifs ~ mass.
Besides, the calculated \Nifs ~ mass is slightly overestimated
since we have adopted 21-isotopes nucleosynthesis (Section \ref{dis:method}).
Moreover, the calculated \Nifs ~ mass is the upper limit (Section \ref{dis1:Ni})
as the entire \Nifs ~ is not necessarily ejected.
Therefore, we conclude that majority of CCSNe need to explode with the timescale of $t_{\rm grow} < 0.3$ sec.
This conclusion is consistent with \citet{Sawada2019},
which suggested “rapid” explosion ($t_{\rm grow} \ltsim$ 0.25 sec) for normal Type II SNe.
Also, our conclusions are in line with those by \citet{Sollerman2021},
which has shown that many SESNe show higher luminosity than that in the models with calibrated-neutrino luminosity.

Note that our calculations are based on one-dimensional artificial explosion,
and much simpler than ab-initio calculations with the neutrino-driven mechanism \citep[e.g.,][]{Janka2012}
and models with calibrated neutrino luminosity \citep[e.g.,][]{Ugliano2012, Ertl2020, Woosley2021}.
Nevertheless, our results give a constraint to the explosion timescale of CCSNe in a model-independent manner:
in order to explain the synthesized \Nifs ~ mass in the majority of the observed SESNe,
any mechanisms which corresponds to the one-dimensional calculation with $t_{\rm grow} < 0.3$ sec is preferable.

It is also noted that even the rapid explosions with 0.1 sec cannot reproduce the \Nifs ~ masses of some SNe.
There are several possibilities to produce additional \Nifs.
If some SNe explode by leaving a NS with small ejecta mass and eject material from a deeper core, more \Nifs ~ can be ejected.
Although such a case is likely to be a minority according to the observed NS mass distribution \citep{Lattimer2012},
our calculations with an inner boundary of 1.2 \Msun ~ show that
additional $\sim$ 0.1 \Msun ~ of \Nifs ~ can be ejected
as compared with the case with the inner boundary of 1.4 \Msun ~ (see Appendix B).
Another mechanism that additionally produces \Nifs ~ is neutrino-driven wind after the explosive synthesis
\citep[e.g., ][]{Wongwathanarat2017, Wanajo2018, Sawada2021},
which can increase \ltsim 0.05 \Msun ~ of \Nifs.
As discussed in Section \ref{dis1:Ni},
it is difficult to explain the luminosity of SNe with $M_{\rm Ni} > 0.2 \  M_{\rm ej}$
by the radioactive decay of \Nifs.
Such SNe may be regarded as a category of SLSNe, which would require other power sources,
such as magnetars \citep[e.g., ][]{Kasen2010, Woosley2010, Mosta2014}.

Constraints on the explosion timescale will be reinforced by future observations.
Although our calculations show the overall trend that the \Nifs ~ mass correlates with the ejecta mass as mentioned in Section \ref{dis1:Ni},
the observational results do not seem to show the clear trend.
This is because of the small sample size for high ejecta mass (\gtsim $5 ~ M_\odot$).
When more SNe with high ejecta mass are discovered in future observations,
we can test the trend with the ejecta mass and the relation between \Nifs ~ production and the progenitor structure.

\section{Summary}
\label{sec:sum}

In this work, we focus on \Nifs ~ to give constraints on the explosion timescale of CCSNe.
We perform systematic analysis of multi-band light curves of 82 SESNe (including SLSNe-I) to obtain bolometric light curves,
which is one of the largest sample numbers among the research regarding bolometric light curves of SESNe
based on multi-band SEDs.
We measure the decline timescale \deltat ~ of the light curves and the peak magnitude \Mpeak.
There is no clear correlation between them, which is consistent with previous studies.
Also, the distributions of these parameters are similar among Type IIb, Ib, and Ic SNe.
Then, we estimate the ejecta mass and the \Nifs ~ mass from the decline timescale and the peak magnitude
using the semi-analytic model \citep{Arnett1982, Chatzopoulos2012}.

We carry out one-dimensional hydrodynamics and nucleosynthesis calculations,
varying the progenitor mass and the explosion timescale to evaluate the \Nifs ~ mass.
We find that the synthesized \Nifs ~ mass is less than $\sim 0.2 $ \Mej,
showing that the maximum luminosity of normal CCSNe powered by \Nifs ~
is determined by  $M_{\rm Ni} \ltsim  0.2 \ M_{\rm ej}$.
This suggests that SNe exceeding the upper limit of the \Nifs ~ mass are likely to have other power sources.

From the comparison of the \Nifs ~ mass estimated from the observations and the calculations,
we find that the explosions with 1.0 sec (slow explosions preferred by ab-initio calculations)
can only explain less than $50 ~ \%$ of the \Nifs ~ mass of the SNe,
and the explosions with 0.3 sec can explain the majority of \Nifs ~ mass of the SNe in our sample.
Therefore, we conclude that the explosion timescale of CCSNe is
shorter than 0.3 sec to explain the \Nifs ~ mass of the majority of SESNe.

\begin{acknowledgements}
This work was supported by JSPS KAKENHI Grant Number JP21J22515 and Graduate Program on Physics for the Universe (GP-PU), Tohoku University (S.S.); JSPS KAKENHI Grant Number JP17H06363, JP19H00694, JP20H00158, JP20H00179, and JP21H04997 (M.T.);  JP21J00825 and JP21K13964 (R.S.); JP18K13585, JP20H00174, JP21K13966, and JP21H04997 (T.M.).
\end{acknowledgements}

\bibliography{references}{}

\begin{thebibliography}{}
\expandafter\ifx\csname natexlab\endcsname\relax\def\natexlab#1{#1}\fi
\providecommand{\url}[1]{\href{#1}{#1}}

\bibitem[{{Afsariardchi} {et~al.}(2021){Afsariardchi}, {Drout}, {Khatami},
  {Matzner}, {Moon}, \& {Ni}}]{Afsariardchi2021}
{Afsariardchi}, N., {Drout}, M.~R., {Khatami}, D.~K., {et~al.} 2021, \apj, 918,
  89

\bibitem[{{Anderson}(2019)}]{Anderson2019}
{Anderson}, J.~P. 2019, \aap, 628, A7

\bibitem[{{Arcavi} {et~al.}(2016){Arcavi}, {Wolf}, {Howell}, {Bildsten}, \&
  {PTF}}]{Arcavi2016}
{Arcavi}, I., {Wolf}, W.~M., {Howell}, D.~A., {Bildsten}, L., \& {PTF}, S.
  2016, in American Astronomical Society Meeting Abstracts, Vol. 227, American
  Astronomical Society Meeting Abstracts \#227, 208.02

\bibitem[{{Arnett}(1966)}]{Arnett1966}
{Arnett}, W.~D. 1966, Canadian Journal of Physics, 44, 2553

\bibitem[{{Arnett}(1979)}]{Arnett1979}
---. 1979, \apjl, 230, L37

\bibitem[{{Arnett}(1982)}]{Arnett1982}
---. 1982, \apj, 253, 785

\bibitem[{{Barbarino} {et~al.}(2021){Barbarino}, {Sollerman}, {Taddia},
  {Fremling}, {Karamehmetoglu}, {Arcavi}, {Gal-Yam}, {Laher}, {Schulze},
  {Wozniak}, \& {Yan}}]{Barbarino2021}
{Barbarino}, C., {Sollerman}, J., {Taddia}, F., {et~al.} 2021, \aap, 651, A81

\bibitem[{{Bethe} \& {Wilson}(1985)}]{Bethe1985}
{Bethe}, H.~A., \& {Wilson}, J.~R. 1985, \apj, 295, 14

\bibitem[{{Blanton} \& {Roweis}(2007)}]{Blanton2007}
{Blanton}, M.~R., \& {Roweis}, S. 2007, \aj, 133, 734

\bibitem[{{Bollig} {et~al.}(2021){Bollig}, {Yadav}, {Kresse}, {Janka},
  {M{\"u}ller}, \& {Heger}}]{Bollig2021}
{Bollig}, R., {Yadav}, N., {Kresse}, D., {et~al.} 2021, \apj, 915, 28

\bibitem[{{Burrows} {et~al.}(2019){Burrows}, {Radice}, \&
  {Vartanyan}}]{Burrows2019}
{Burrows}, A., {Radice}, D., \& {Vartanyan}, D. 2019, \mnras, 485, 3153

\bibitem[{{Burrows} {et~al.}(2020){Burrows}, {Radice}, {Vartanyan}, {Nagakura},
  {Skinner}, \& {Dolence}}]{Burrows2020}
{Burrows}, A., {Radice}, D., {Vartanyan}, D., {et~al.} 2020, \mnras, 491, 2715

\bibitem[{{Burrows} \& {Vartanyan}(2021)}]{Burrows2021}
{Burrows}, A., \& {Vartanyan}, D. 2021, \nat, 589, 29

\bibitem[{{Cano}(2013)}]{Cano2013}
{Cano}, Z. 2013, \mnras, 434, 1098

\bibitem[{{Chatzopoulos} {et~al.}(2012){Chatzopoulos}, {Wheeler}, \&
  {Vinko}}]{Chatzopoulos2012}
{Chatzopoulos}, E., {Wheeler}, J.~C., \& {Vinko}, J. 2012, \apj, 746, 121

\bibitem[{{Chevalier}(1992)}]{Chevalier1992}
{Chevalier}, R.~A. 1992, \apj, 394, 599

\bibitem[{{Colgate} \& {White}(1966)}]{Colgate1966}
{Colgate}, S.~A., \& {White}, R.~H. 1966, \apj, 143, 626

\bibitem[{{De Cia} {et~al.}(2018){De Cia}, {Gal-Yam}, {Rubin}, {Leloudas},
  {Vreeswijk}, {Perley}, {Quimby}, {Yan}, {Sullivan}, {Fl{\"o}rs}, {Sollerman},
  {Bersier}, {Cenko}, {Gal-Yam}, {Maguire}, {Ofek}, {Prentice}, {Schulze},
  {Spyromilio}, {Valenti}, {Arcavi}, {Corsi}, {Howell}, {Mazzali}, {Kasliwal},
  {Taddia}, \& {Yaron}}]{DeCia2018}
{De Cia}, A., {Gal-Yam}, A., {Rubin}, A., {et~al.} 2018, \apj, 860, 100

\bibitem[{{Dessart} {et~al.}(2016){Dessart}, {Hillier}, {Woosley}, {Livne},
  {Waldman}, {Yoon}, \& {Langer}}]{Dessart2016}
{Dessart}, L., {Hillier}, D.~J., {Woosley}, S., {et~al.} 2016, \mnras, 458,
  1618

\bibitem[{{Drout} {et~al.}(2011){Drout}, {Soderberg}, {Gal-Yam}, {Cenko},
  {Fox}, {Leonard}, {Sand}, {Moon}, {Arcavi}, \& {Green}}]{Drout2011}
{Drout}, M.~R., {Soderberg}, A.~M., {Gal-Yam}, A., {et~al.} 2011, \apj, 741, 97

\bibitem[{{Eldridge} {et~al.}(2008){Eldridge}, {Izzard}, \&
  {Tout}}]{Eldridge2008}
{Eldridge}, J.~J., {Izzard}, R.~G., \& {Tout}, C.~A. 2008, \mnras, 384, 1109

\bibitem[{{Ertl} {et~al.}(2020){Ertl}, {Woosley}, {Sukhbold}, \&
  {Janka}}]{Ertl2020}
{Ertl}, T., {Woosley}, S.~E., {Sukhbold}, T., \& {Janka}, H.~T. 2020, \apj,
  890, 51

\bibitem[{{Gal-Yam}(2012)}]{GalYam2012}
{Gal-Yam}, A. 2012, Science, 337, 927

\bibitem[{{Guillochon} {et~al.}(2017){Guillochon}, {Parrent}, {Kelley}, \&
  {Margutti}}]{guillochon2017jan}
{Guillochon}, J., {Parrent}, J., {Kelley}, L.~Z., \& {Margutti}, R. 2017, \apj,
  835, 64

\bibitem[{{Hamuy}(2003)}]{Hamuy2003}
{Hamuy}, M. 2003, \apj, 582, 905

\bibitem[{{Janka}(2012)}]{Janka2012}
{Janka}, H.-T. 2012, Annual Review of Nuclear and Particle Science, 62, 407

\bibitem[{{Kasen} \& {Bildsten}(2010)}]{Kasen2010}
{Kasen}, D., \& {Bildsten}, L. 2010, \apj, 717, 245

\bibitem[{{Khatami} \& {Kasen}(2019)}]{Khatami2019}
{Khatami}, D.~K., \& {Kasen}, D.~N. 2019, \apj, 878, 56

\bibitem[{{Lattimer}(2012)}]{Lattimer2012}
{Lattimer}, J.~M. 2012, \Annual, 62, 485

\bibitem[{{Lentz} {et~al.}(2015){Lentz}, {Bruenn}, {Hix}, {Mezzacappa},
  {Messer}, {Endeve}, {Blondin}, {Harris}, {Marronetti}, \&
  {Yakunin}}]{Lentz2015}
{Lentz}, E.~J., {Bruenn}, S.~W., {Hix}, W.~R., {et~al.} 2015, \apjl, 807, L31

\bibitem[{{Lyman} {et~al.}(2016){Lyman}, {Bersier}, {James}, {Mazzali},
  {Eldridge}, {Fraser}, \& {Pian}}]{Lyman2016}
{Lyman}, J.~D., {Bersier}, D., {James}, P.~A., {et~al.} 2016, \mnras, 457, 328

\bibitem[{{MacFadyen} \& {Woosley}(1999)}]{MacFadyen1999}
{MacFadyen}, A.~I., \& {Woosley}, S.~E. 1999, \apj, 524, 262

\bibitem[{{Maeda} \& {Tominaga}(2009)}]{Maeda2009}
{Maeda}, K., \& {Tominaga}, N. 2009, \mnras, 394, 1317

\bibitem[{{Maund} {et~al.}(2004){Maund}, {Smartt}, {Kudritzki},
  {Podsiadlowski}, \& {Gilmore}}]{Maund2004}
{Maund}, J.~R., {Smartt}, S.~J., {Kudritzki}, R.~P., {Podsiadlowski}, P., \&
  {Gilmore}, G.~F. 2004, \nat, 427, 129

\bibitem[{{Melson} {et~al.}(2015{\natexlab{a}}){Melson}, {Janka}, {Bollig},
  {Hanke}, {Marek}, \& {M{\"u}ller}}]{Melson2015b}
{Melson}, T., {Janka}, H.-T., {Bollig}, R., {et~al.} 2015{\natexlab{a}}, \apjl,
  808, L42

\bibitem[{{Melson} {et~al.}(2015{\natexlab{b}}){Melson}, {Janka}, \&
  {Marek}}]{Melson2015a}
{Melson}, T., {Janka}, H.-T., \& {Marek}, A. 2015{\natexlab{b}}, \apjl, 801,
  L24

\bibitem[{{Meza} \& {Anderson}(2020)}]{Meza2020}
{Meza}, N., \& {Anderson}, J.~P. 2020, \aap, 641, A177

\bibitem[{{Modjaz} {et~al.}(2009){Modjaz}, {Li}, {Butler}, {Chornock},
  {Perley}, {Blondin}, {Bloom}, {Filippenko}, {Kirshner}, {Kocevski},
  {Poznanski}, {Hicken}, {Foley}, {Stringfellow}, {Berlind}, {Barrado y
  Navascues}, {Blake}, {Bouy}, {Brown}, {Challis}, {Chen}, {de Vries},
  {Dufour}, {Falco}, {Friedman}, {Ganeshalingam}, {Garnavich}, {Holden},
  {Illingworth}, {Lee}, {Liebert}, {Marion}, {Olivier}, {Prochaska},
  {Silverman}, {Smith}, {Starr}, {Steele}, {Stockton}, {Williams}, \&
  {Wood-Vasey}}]{Modjaz2009}
{Modjaz}, M., {Li}, W., {Butler}, N., {et~al.} 2009, \apj, 702, 226

\bibitem[{{Moriya} {et~al.}(2018){Moriya}, {Sorokina}, \&
  {Chevalier}}]{Moriya2018a}
{Moriya}, T.~J., {Sorokina}, E.~I., \& {Chevalier}, R.~A. 2018, \ssr, 214, 59

\bibitem[{{Morozova} {et~al.}(2015){Morozova}, {Ott}, \& {Piro}}]{Morozova2015}
{Morozova}, V., {Ott}, C.~D., \& {Piro}, A.~L. 2015, {SNEC: SuperNova Explosion
  Code}, , , ascl:1505.033

\bibitem[{{M{\"o}sta} {et~al.}(2014){M{\"o}sta}, {Richers}, {Ott}, {Haas},
  {Piro}, {Boydstun}, {Abdikamalov}, {Reisswig}, \& {Schnetter}}]{Mosta2014}
{M{\"o}sta}, P., {Richers}, S., {Ott}, C.~D., {et~al.} 2014, \apjl, 785, L29

\bibitem[{{M{\"u}ller} {et~al.}(2017){M{\"u}ller}, {Melson}, {Heger}, \&
  {Janka}}]{Muller2017}
{M{\"u}ller}, B., {Melson}, T., {Heger}, A., \& {Janka}, H.-T. 2017, \mnras,
  472, 491

\bibitem[{{M{\"u}ller} {et~al.}(2019){M{\"u}ller}, {Tauris}, {Heger},
  {Banerjee}, {Qian}, {Powell}, {Chan}, {Gay}, \& {Langer}}]{Muller2019}
{M{\"u}ller}, B., {Tauris}, T.~M., {Heger}, A., {et~al.} 2019, \mnras, 484,
  3307

\bibitem[{{Nadyozhin}(1994)}]{Nadyozhin1994}
{Nadyozhin}, D.~K. 1994, \apjs, 92, 527

\bibitem[{{Nicholl}(2021)}]{Nicholl2021}
{Nicholl}, M. 2021, arXiv e-prints, arXiv:2109.08697

\bibitem[{{O'Connor} \& {Ott}(2011)}]{OConnor2011}
{O'Connor}, E., \& {Ott}, C.~D. 2011, \apj, 730, 70

\bibitem[{{Ott} {et~al.}(2018){Ott}, {Roberts}, {da Silva Schneider}, {Fedrow},
  {Haas}, \& {Schnetter}}]{Ott2018}
{Ott}, C.~D., {Roberts}, L.~F., {da Silva Schneider}, A., {et~al.} 2018, \apjl,
  855, L3

\bibitem[{{Ouchi} {et~al.}(2021){Ouchi}, {Maeda}, {Anderson}, \&
  {Sawada}}]{Ouchi2021}
{Ouchi}, R., {Maeda}, K., {Anderson}, J.~P., \& {Sawada}, R. 2021, arXiv
  e-prints, arXiv:2109.00603

\bibitem[{{Pejcha} \& {Prieto}(2015)}]{Pejcha2015}
{Pejcha}, O., \& {Prieto}, J.~L. 2015, \apj, 806, 225

\bibitem[{{Perley} {et~al.}(2020){Perley}, {Fremling}, {Sollerman}, {Miller},
  {Dahiwale}, {Sharma}, {Bellm}, {Biswas}, {Brink}, {Bruch}, {De}, {Dekany},
  {Drake}, {Duev}, {Filippenko}, {Gal-Yam}, {Goobar}, {Graham}, {Graham}, {Ho},
  {Irani}, {Kasliwal}, {Kim}, {Kulkarni}, {Mahabal}, {Masci}, {Modak}, {Neill},
  {Nordin}, {Riddle}, {Soumagnac}, {Strotjohann}, {Schulze}, {Taggart},
  {Tzanidakis}, {Walters}, \& {Yan}}]{Perley2020}
{Perley}, D.~A., {Fremling}, C., {Sollerman}, J., {et~al.} 2020, \apj, 904, 35

\bibitem[{{Prentice} {et~al.}(2021){Prentice}, {Inserra}, {Schulze}, {Nicholl},
  {Mazzali}, {Vergani}, {Galbany}, {Anderson}, {Ashall}, {Chen}, {Deckers},
  {Manche no}, {Gonz{\'a}lez D{\'\i}az}, {Gonz{\'a}lez-Gait{\'a}n},
  {Gromadzki}, {Guti{\'e}rrez}, {Harvey}, {Kozyreva}, {Magee}, {Maguire},
  {M{\"u}ller-Bravo}, {Mu{\~n}oz Torres}, {Pessi}, {Sollerman}, {Teffs},
  {Terwel}, \& {Young}}]{Prentice2021}
{Prentice}, S.~J., {Inserra}, C., {Schulze}, S., {et~al.} 2021, \mnras,
  arXiv:2109.14572

\bibitem[{{Sana} {et~al.}(2012){Sana}, {de Mink}, {de Koter}, {Langer},
  {Evans}, {Gieles}, {Gosset}, {Izzard}, {Le Bouquin}, \&
  {Schneider}}]{Sana2012}
{Sana}, H., {de Mink}, S.~E., {de Koter}, A., {et~al.} 2012, Science, 337, 444

\bibitem[{{Sawada} \& {Maeda}(2019)}]{Sawada2019}
{Sawada}, R., \& {Maeda}, K. 2019, \apj, 886, 47

\bibitem[{{Sawada} \& {Suwa}(2021)}]{Sawada2021}
{Sawada}, R., \& {Suwa}, Y. 2021, \apj, 908, 6

\bibitem[{{Schlegel} {et~al.}(1998){Schlegel}, {Finkbeiner}, \&
  {Davis}}]{Schlegel1998}
{Schlegel}, D.~J., {Finkbeiner}, D.~P., \& {Davis}, M. 1998, \apj, 500, 525

\bibitem[{{Seitenzahl} {et~al.}(2009){Seitenzahl}, {Townsley}, {Peng}, \&
  {Truran}}]{Seitenzahl2009}
{Seitenzahl}, I.~R., {Townsley}, D.~M., {Peng}, F., \& {Truran}, J.~W. 2009,
  Atomic Data and Nuclear Data Tables, 95, 96

\bibitem[{{Sollerman} {et~al.}(2021){Sollerman}, {Yang}, {Perley}, {Schulze},
  {Fremling}, {Kasliwal}, {Shin}, \& {Racine}}]{Sollerman2021}
{Sollerman}, J., {Yang}, S., {Perley}, D., {et~al.} 2021, arXiv e-prints,
  arXiv:2109.14339

\bibitem[{{Sravan} {et~al.}(2019){Sravan}, {Marchant}, \&
  {Kalogera}}]{Sravan2019}
{Sravan}, N., {Marchant}, P., \& {Kalogera}, V. 2019, \apj, 885, 130

\bibitem[{{Stockinger} {et~al.}(2020){Stockinger}, {Janka}, {Kresse}, {Melson},
  {Ertl}, {Gabler}, {Gessner}, {Wongwathanarat}, {Tolstov}, {Leung}, {Nomoto},
  \& {Heger}}]{Stockinger2020}
{Stockinger}, G., {Janka}, H.~T., {Kresse}, D., {et~al.} 2020, \mnras, 496,
  2039

\bibitem[{{Stritzinger} {et~al.}(2018){Stritzinger}, {Taddia}, {Burns},
  {Phillips}, {Bersten}, {Contreras}, {Folatelli}, {Holmbo}, {Hsiao},
  {Hoeflich}, {Leloudas}, {Morrell}, {Sollerman}, \&
  {Suntzeff}}]{Stritzinger2018}
{Stritzinger}, M.~D., {Taddia}, F., {Burns}, C.~R., {et~al.} 2018, \aap, 609,
  A135

\bibitem[{{Sukhbold} {et~al.}(2016){Sukhbold}, {Ertl}, {Woosley}, {Brown}, \&
  {Janka}}]{Sukhbold2016}
{Sukhbold}, T., {Ertl}, T., {Woosley}, S.~E., {Brown}, J.~M., \& {Janka}, H.~T.
  2016, \apj, 821, 38

\bibitem[{{Summa} {et~al.}(2018){Summa}, {Janka}, {Melson}, \&
  {Marek}}]{Summa2018}
{Summa}, A., {Janka}, H.-T., {Melson}, T., \& {Marek}, A. 2018, \apj, 852, 28

\bibitem[{{Sutherland} \& {Wheeler}(1984)}]{Sutherland1984}
{Sutherland}, P.~G., \& {Wheeler}, J.~C. 1984, \apj, 280, 282

\bibitem[{{Suwa} \& {Tominaga}(2015)}]{Suwa2015}
{Suwa}, Y., \& {Tominaga}, N. 2015, \mnras, 451, 282

\bibitem[{{Suwa} {et~al.}(2019){Suwa}, {Tominaga}, \& {Maeda}}]{Suwa2019}
{Suwa}, Y., {Tominaga}, N., \& {Maeda}, K. 2019, \mnras, 483, 3607

\bibitem[{{Taddia} {et~al.}(2015){Taddia}, {Sollerman}, {Leloudas},
  {Stritzinger}, {Valenti}, {Galbany}, {Kessler}, {Schneider}, \&
  {Wheeler}}]{Taddia2015}
{Taddia}, F., {Sollerman}, J., {Leloudas}, G., {et~al.} 2015, \aap, 574, A60

\bibitem[{{Taddia} {et~al.}(2018){Taddia}, {Stritzinger}, {Bersten}, {Baron},
  {Burns}, {Contreras}, {Holmbo}, {Hsiao}, {Morrell}, {Phillips}, {Sollerman},
  \& {Suntzeff}}]{Taddia2018}
{Taddia}, F., {Stritzinger}, M.~D., {Bersten}, M., {et~al.} 2018, \aap, 609,
  A136

\bibitem[{{Takiwaki} {et~al.}(2014){Takiwaki}, {Kotake}, \&
  {Suwa}}]{Takiwaki2014}
{Takiwaki}, T., {Kotake}, K., \& {Suwa}, Y. 2014, \apj, 786, 83

\bibitem[{{Timmes}(1999)}]{Timmes1999}
{Timmes}, F.~X. 1999, \apjs, 124, 241

\bibitem[{{Tully} \& {Fisher}(1977)}]{Tully1977}
{Tully}, R.~B., \& {Fisher}, J.~R. 1977, \aap, 500, 105

\bibitem[{{Ugliano} {et~al.}(2012){Ugliano}, {Janka}, {Marek}, \&
  {Arcones}}]{Ugliano2012}
{Ugliano}, M., {Janka}, H.-T., {Marek}, A., \& {Arcones}, A. 2012, \apj, 757,
  69

\bibitem[{{Vartanyan} {et~al.}(2021){Vartanyan}, {Laplace}, {Renzo},
  {G{\"o}tberg}, {Burrows}, \& {de Mink}}]{Vartanyan2021}
{Vartanyan}, D., {Laplace}, E., {Renzo}, M., {et~al.} 2021, arXiv e-prints,
  arXiv:2104.03317

\bibitem[{{Wanajo} {et~al.}(2018){Wanajo}, {M{\"u}ller}, {Janka}, \&
  {Heger}}]{Wanajo2018}
{Wanajo}, S., {M{\"u}ller}, B., {Janka}, H.-T., \& {Heger}, A. 2018, \apj, 852,
  40

\bibitem[{{Wongwathanarat} {et~al.}(2017){Wongwathanarat}, {Janka},
  {M{\"u}ller}, {Pllumbi}, \& {Wanajo}}]{Wongwathanarat2017}
{Wongwathanarat}, A., {Janka}, H.-T., {M{\"u}ller}, E., {Pllumbi}, E., \&
  {Wanajo}, S. 2017, \apj, 842, 13

\bibitem[{{Woosley}(1988)}]{Woosley1988}
{Woosley}, S.~E. 1988, \apj, 330, 218

\bibitem[{{Woosley}(2010)}]{Woosley2010}
---. 2010, \apjl, 719, L204

\bibitem[{{Woosley} {et~al.}(2002){Woosley}, {Heger}, \&
  {Weaver}}]{Woosley2002}
{Woosley}, S.~E., {Heger}, A., \& {Weaver}, T.~A. 2002, Reviews of Modern
  Physics, 74, 1015

\bibitem[{{Woosley} {et~al.}(2021){Woosley}, {Sukhbold}, \&
  {Kasen}}]{Woosley2021}
{Woosley}, S.~E., {Sukhbold}, T., \& {Kasen}, D.~N. 2021, \apj, 913, 145

\bibitem[{{Woosley} \& {Weaver}(1995)}]{Woosley1995}
{Woosley}, S.~E., \& {Weaver}, T.~A. 1995, \apjs, 101, 181

\bibitem[{{Yoon} {et~al.}(2010){Yoon}, {Woosley}, \& {Langer}}]{Yoon2010}
{Yoon}, S.~C., {Woosley}, S.~E., \& {Langer}, N. 2010, \apj, 725, 940

\bibitem[{{Zheng} {et~al.}(2022){Zheng}, {Stahl}, {de Jaeger}, {Filippenko},
  {Wang}, {Gan}, {Brink}, {Altunin}, {Baer-Way}, {Bigley}, {Blanchard},
  {Blanchard}, {Bradley}, {Cargill}, {Casper}, {Chapman}, {Chander}, {Channa},
  {Choi}, {Choksi}, {Chu}, {Clubb}, {Cohen}, {Dalba}, {deGraw}, {de
  Kouchkovsky}, {Ellison}, {Falcon}, {Fox}, {Fuller}, {Ganeshalingam},
  {Girish}, {Gould}, {Halevi}, {Halle}, {Hayakawa}, {Hardy}, {Hestenes},
  {Hoffman}, {Hyland}, {Jeffers}, {Jennings}, {Kandrashoff}, {Khodanian},
  {Kim}, {Kim}, {Kislak}, {Krishnan}, {Kumar}, {Kumar}, {Leja}, {Leonard},
  {Li}, {Li}, {Lian}, {Liu}, {Lowe}, {Lu}, {Ma}, {Mason}, {May}, {McAllister},
  {McGinness}, {Modak}, {Molloy}, {Murakami}, {Nayak}, {Perera}, {Pina},
  {Punjabi}, {Rikhter}, {Ross}, {Sipple}, {Soler}, {Stegman}, {Stephens},
  {Sunseri}, {Tang}, {Taylor}, {Thrasher}, {Van Dyk}, {Wang}, {Wayland},
  {Wilkins}, {Yagubyan}, {Yuk}, {Yunus}, \& {Zhang}}]{Zheng2022}
{Zheng}, W., {Stahl}, B.~E., {de Jaeger}, T., {et~al.} 2022, \mnras,
  arXiv:2203.05596

\end{thebibliography}
\bibliographystyle{aasjournal}

\section*{Appendix A: \Mej ~ vs. \MNi ~ from a constant kinetic energy}
\label{appA}

\begin{figure*}[ht]
\begin{center}
  \begin{tabular}{c}
    \begin{minipage}{0.5\hsize}
      \begin{center}
        \includegraphics[width= 0.97\linewidth, bb=7 7 460 580  ]{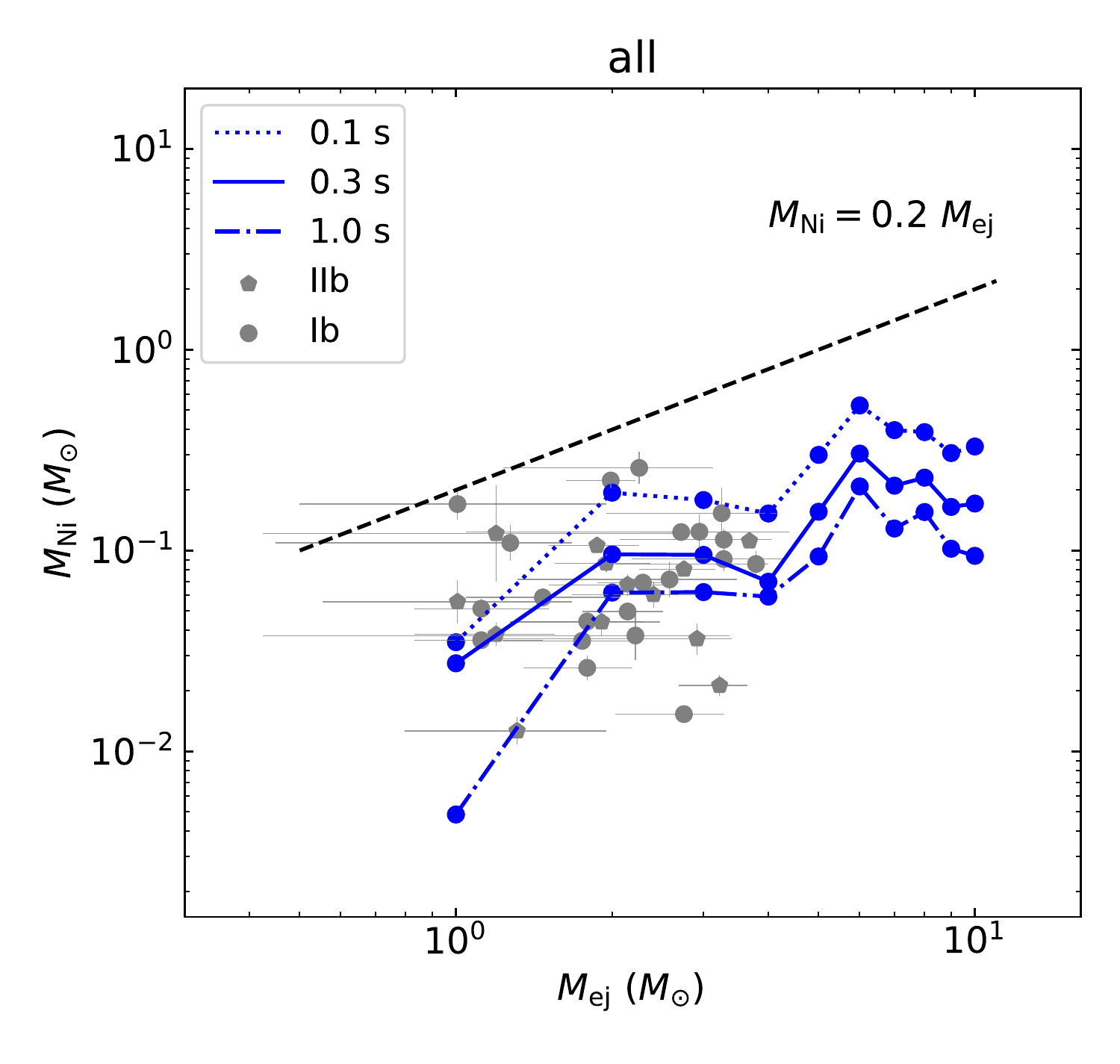}
        \end{center}
    \end{minipage}    
     \begin{minipage}{0.5\hsize}
      \begin{center}
       \includegraphics[width= 0.97\linewidth, bb=7 7 460 580  ]{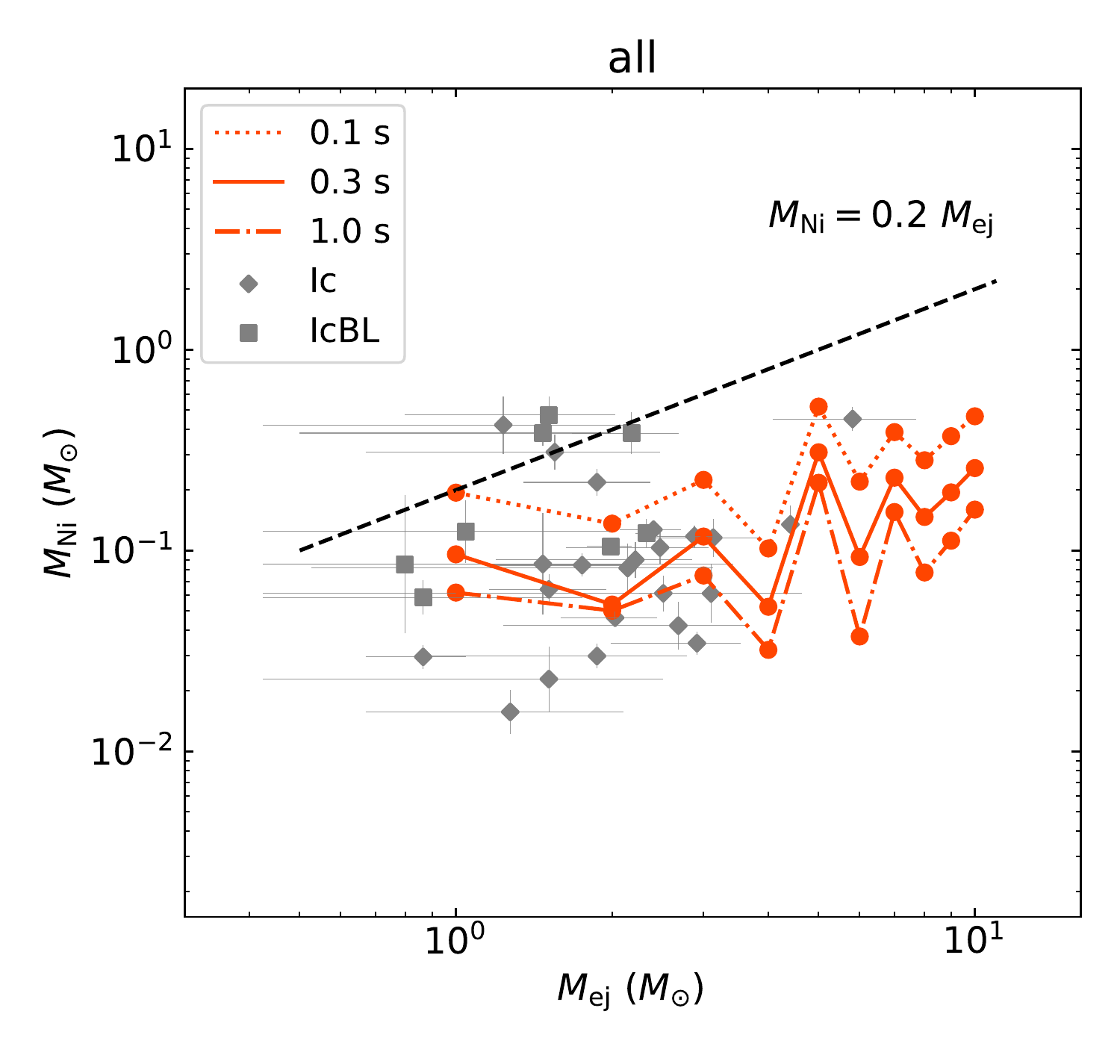}
       \end{center}
    \end{minipage}    
  \end{tabular}  
\caption{Same as Figure \ref{fig:cmpr} but the kinetic energy used for the estimate of the ejecta mass
and used in the calculations are fixed to $E_{\rm k} = 10^{51}$ erg.}
  \label{fig:cmpra}
  \end{center}
\end{figure*}

In this Appendix, we show results of parameter estimation and nucleosynthesis 
by adopting a constant kinetic energy $E_{\rm k} = 10^{51}$ erg.
Figure \ref{fig:cmpra} represents a relation between the ejecta mass and the \Nifs ~ mass.
The ejecta masses estimated from calculations are concentrated on $\sim$ 1 \Msun ~ to $\sim$ 2 \Msun ~
in contrast to the case that the kinetic energy is proportional to the ejecta mass
with the constant velocity $v_{\rm ej} = 10,000 {\rm ~ km ~ s^{-1}}$ (Figure \ref{fig:cmpr}).
This is because the ejecta velocity is observationally estimated to concentrate
around $v_{\rm ej} = 10,000 {\rm ~ km ~ s^{-1}}$  \citep[e.g., ][]{Lyman2016},
and then the typical ejecta mass becomes $\sim$ (5/3) \Msun ~
from the relation of $E_{\rm k} = (3/10) {M_{\rm ej}} {v_{\rm ej}}^2$.
We here emphasize that the case that the kinetic energy is proportional to the ejecta mass assumed in the main text
is more realistic as shown in previous studies \citep[e.g.,][]{Taddia2018}.
All the \Nifs ~ masses except for the case of \Mej ~$= 1$ \Msun ~
are lower than those in the case of the constant ejecta velocity (Figure \ref{fig:cmpr})
because the kinetic energy $E_{\rm k} = 10^{51}$ erg is higher only when the ejecta mass is 1 \Msun.
The fraction that the \Nifs ~ mass estimated from the observations can be explained
by the calculations with each explosion timescale
is almost same as the case of the kinetic energy proportional to the ejecta mass.
This does not affect our conclusions on the explosion timescale of CCSNe.

\section*{Appendix B: inner boundary}
\label{appB}

\begin{figure*}[ht]
\begin{center}
  \begin{tabular}{c}
    \begin{minipage}{0.5\hsize}
      \begin{center}
        \includegraphics[width= 0.65 \linewidth, bb=120 0 500 560]{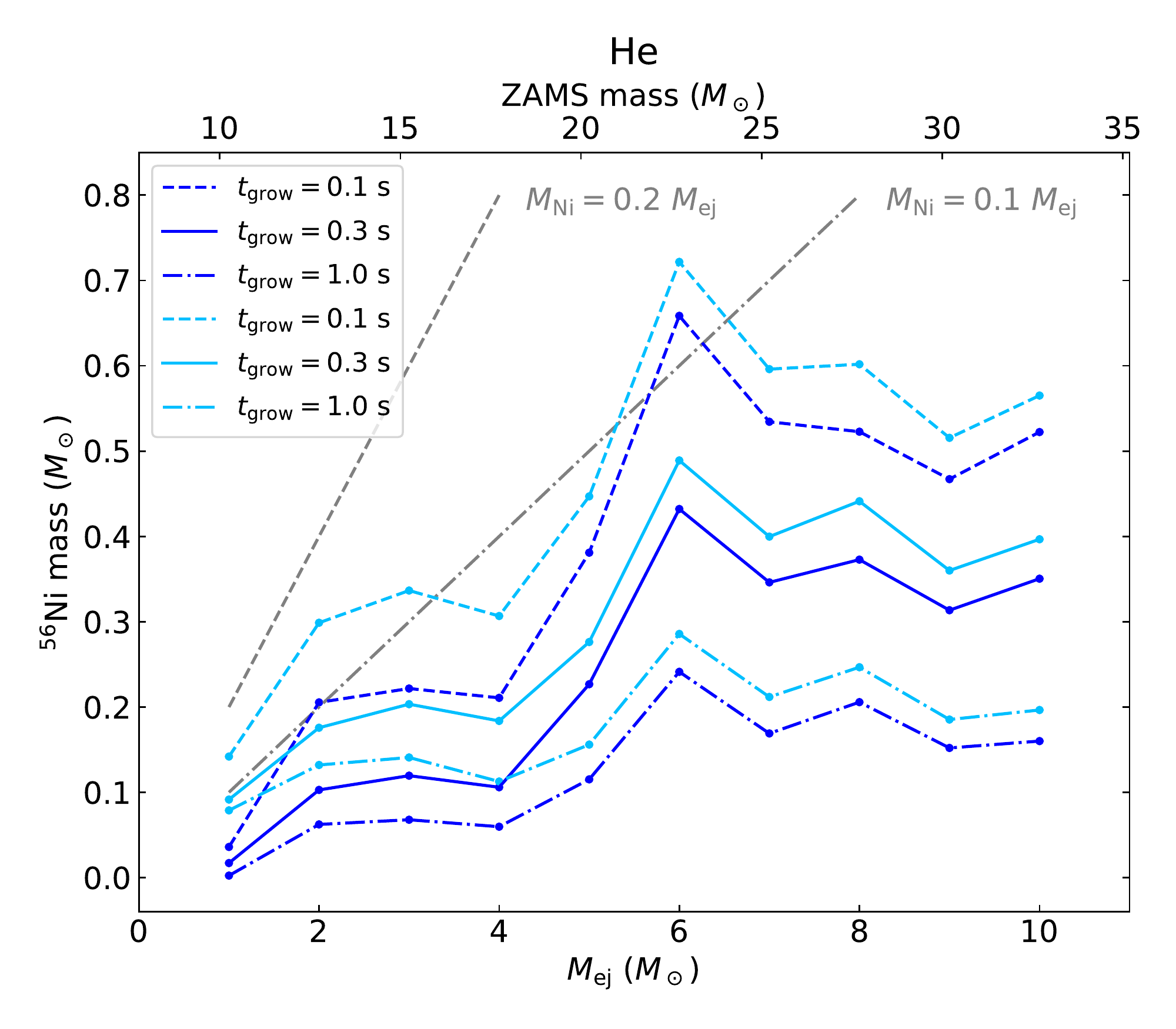}
        \end{center}
    \end{minipage}    
     \begin{minipage}{0.5\hsize}
      \begin{center}
       \includegraphics[width= 0.65 \linewidth, bb=120 0 500 560]{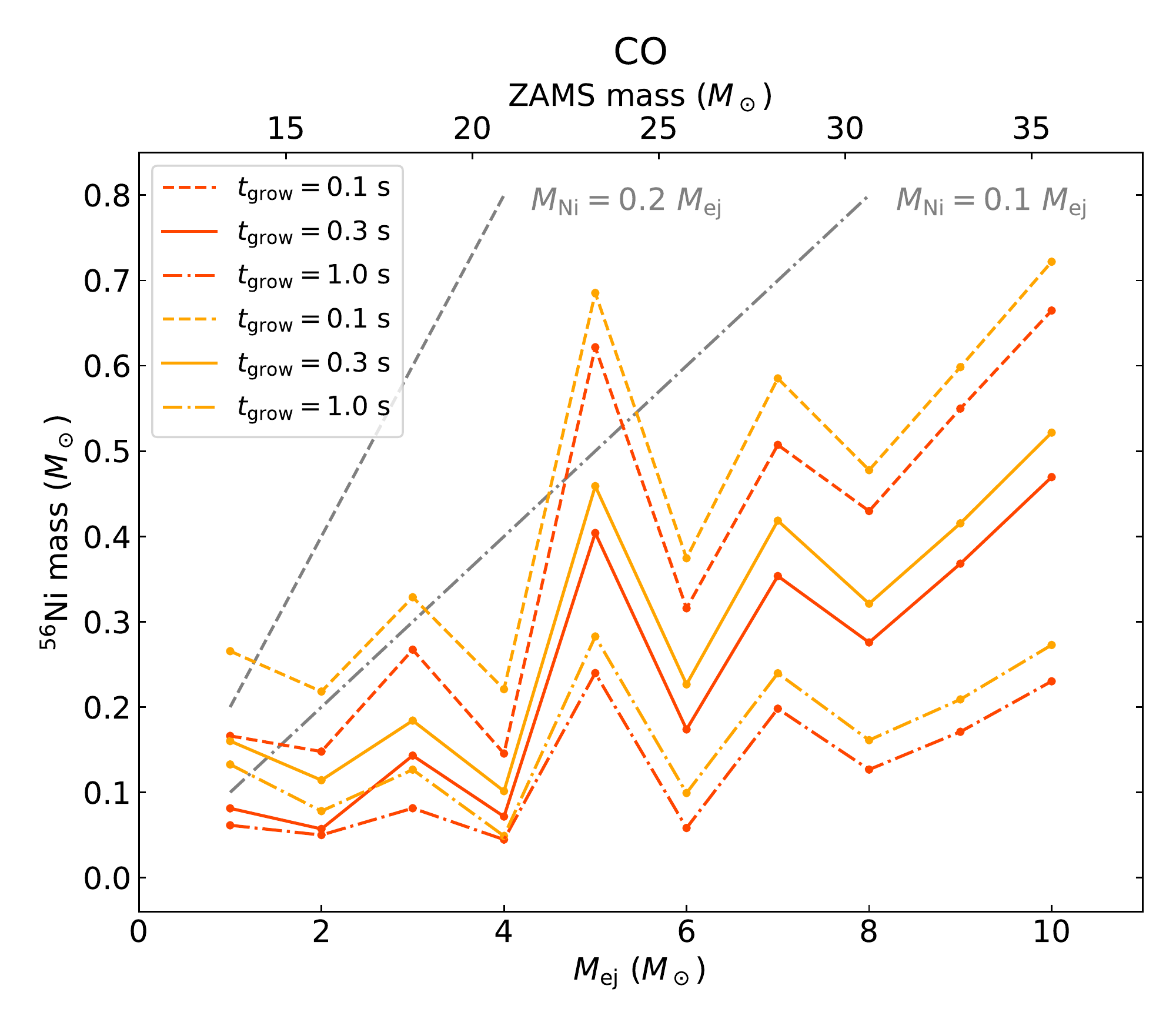}
       \end{center}
    \end{minipage}    
  \end{tabular}  
  \caption{A relation between the ejecta mass and the \Nifs ~ mass.
  The light blue lines and the orange lines are the results of the inner boundary of 1.2 \Msun ~ in the left and right panels, respectively.}
  \label{fig:nib}
  \end{center}
\end{figure*}

Here, we show the results of the calculations by changing the inner boundary of the calculations
from 1.4 \Msun ~ to 1.2 \Msun.
In the calculations, we just change the inner boundary where the energy is input,
keeping the progenitor model and the injection energy the same.
Table \ref{table:MNib} summarizes the synthesized \Nifs~ mass.
The \Nifs ~ masses with the boundary of 1.2 \Msun ~
are $\sim 0.05 - 0.1$ \Msun ~ higher than the cases with 1.4 \Msun ~ as shown in Figure \ref{fig:nib}.
Not all the material enclosed in $1.2 - 1.4$ \Msun ~ in a mass coordinate becomes \Nifs~ 
because a large part of the innermost material becomes $^{56}$Fe due to the low electron fraction ($Y_{\rm e} \sim 0.46$).

\begin{figure*}[ht]
\begin{center}
  \begin{tabular}{c}
    \begin{minipage}{0.5\hsize}
      \begin{center}
        \includegraphics[width= 0.97\linewidth, bb=7 7 460 580  ]{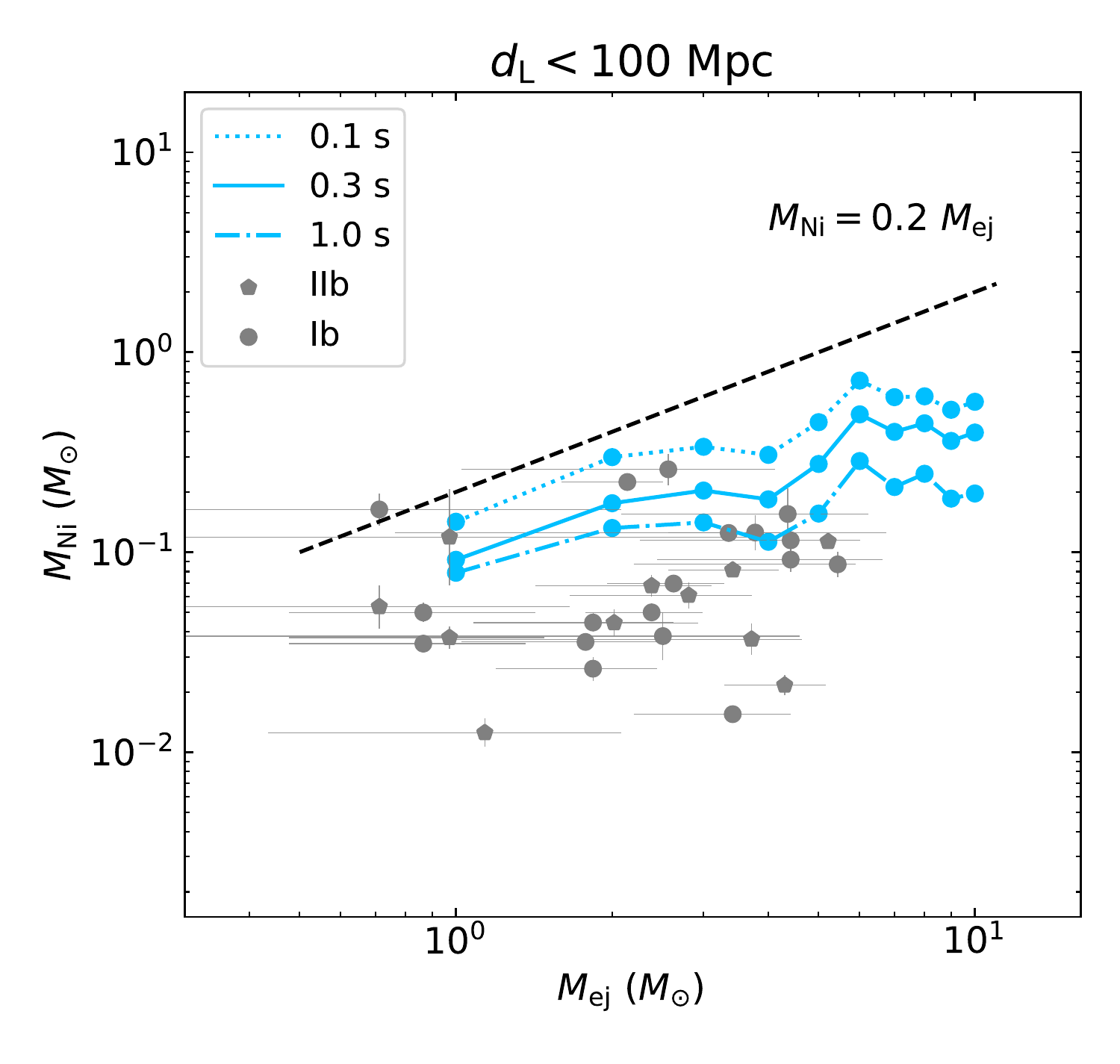}
        \end{center}
    \end{minipage}    
     \begin{minipage}{0.5\hsize}
      \begin{center}
       \includegraphics[width= 0.97\linewidth, bb=7 7 460 580  ]{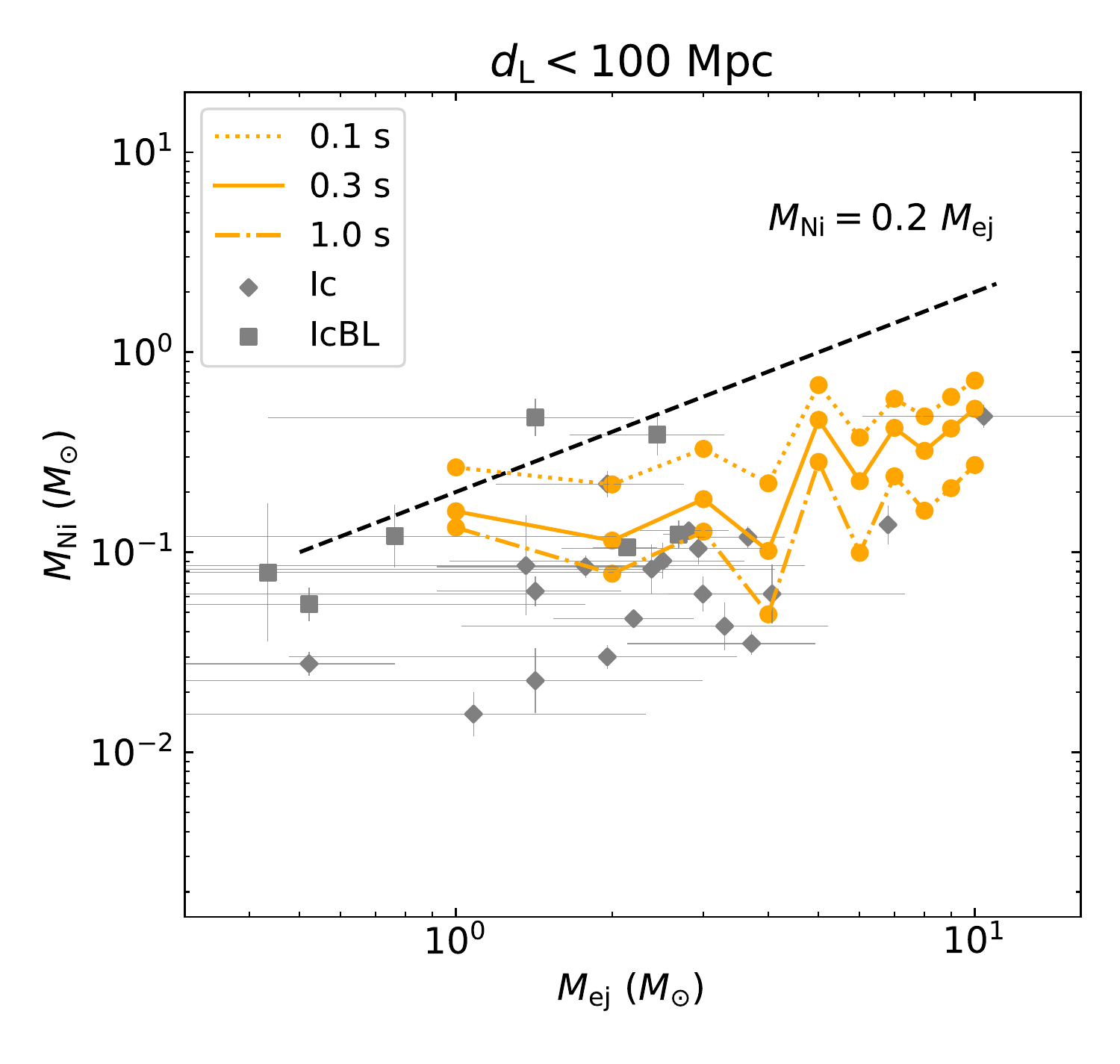}
       \end{center}
    \end{minipage}    
  \end{tabular}  
  \caption{Same as Figure \ref{fig:cmpr} but the inner boundary in the calculations is changed to 1.2 \Msun.}
  \label{fig:cmpr_vlimb}
  \end{center}
\end{figure*}

Figure \ref{fig:cmpr_vlimb} shows the comparison of the results of the inner boundary of  1.2 \Msun ~
with the volume-limited sample of the observations.
The models with the inner boundary 1.2 \Msun ~ can explain the \Nifs ~ mass of some SNe
that cannot be explained by the models with the inner boundary 1.4 \Msun.
This suggests that SNe with higher \Nifs ~ masses may leave NSs with smaller masses.
It is, however, noted that such an explosion is likely to be a minority 
since many of NSs are known from observations to have $\sim 1.4$ \Msun ~ \citep{Lattimer2012}.

\begin{table*}[t]
  \caption{Summary of the synthesized \Nifs ~ masses (in \Msun) for each progenitor model and explosion timescale, $t_{\rm grow}$.}
  \label{table:MNib}
  \centering
  \begin{tabular}{l|ccccc}
    \hline \hline
      &&& {$t_{\rm grow}$ (s)} && \\
Model & 0.01 & 0.03 & 0.1 & 0.3 & 1.0 \\
\hline
He 1 & 0.179 & 0.176 & 0.142 & 0.092 & 0.079 \\ 
He 2 & 0.339 & 0.328 & 0.299 & 0.176 & 0.132 \\ 
He 3 & 0.360 & 0.355 & 0.337 & 0.203 & 0.141 \\ 
He 4 & 0.325 & 0.318 & 0.307 & 0.184 & 0.113 \\ 
He 5 & 0.482 & 0.468 & 0.447 & 0.276 & 0.156 \\ 
He 6 & 0.761 & 0.746 & 0.722 & 0.489 & 0.286 \\ 
He 7 & 0.626 & 0.617 & 0.596 & 0.400 & 0.212 \\ 
He 8 & 0.613 & 0.613 & 0.602 & 0.441 & 0.247 \\ 
He 9 & 0.530 & 0.528 & 0.516 & 0.360 & 0.185 \\ 
He 10 & 0.581 & 0.578 & 0.565 & 0.397 & 0.197 \\ 
\hline 
CO 1 & 0.315 & 0.306 & 0.266 & 0.160 & 0.133 \\ 
CO 2 & 0.260 & 0.252 & 0.218 & 0.114 & 0.078 \\ 
CO 3 & 0.367 & 0.355 & 0.329 & 0.184 & 0.127 \\ 
CO 4 & 0.238 & 0.235 & 0.221 & 0.101 & 0.049 \\ 
CO 5 & 0.728 & 0.710 & 0.685 & 0.459 & 0.282 \\ 
CO 6 & 0.399 & 0.391 & 0.374 & 0.227 & 0.099 \\ 
CO 7 & 0.597 & 0.595 & 0.585 & 0.418 & 0.240 \\ 
CO 8 & 0.493 & 0.490 & 0.478 & 0.321 & 0.161 \\ 
CO 9 & 0.620 & 0.614 & 0.598 & 0.415 & 0.209 \\ 
CO 10 & 0.751 & 0.741 & 0.722 & 0.522 & 0.273 \\ 
\hline 
  \end{tabular}
\end{table*}

\section*{appendix C}
\label{appc}

We summarize information and derived physical parameters for all the SNe in our final sample
in Table \ref{tab:summary}.

\startlongtable
\begin{deluxetable*}{llrcccc}

\tablecaption{A summary of properties of all the samples.} 
\tablehead{ Name & Type  & $d_{\rm L}$ & \deltat & \Mpeak &  \Mej   &   \MNi  \\ 
      &       &     (Mpc)    &  (days)  &  (mag)  & (\Msun) & (\Msun)  }
\startdata
SN 1962L & Ic & 15.2 & $11.2^{+11.2}_{-11.2}$  & $-17.66^{+0.63}_{-0.63}$ & $1.36^{+3.34}_{-1.31}$ & $0.12^{+0.09}_{-0.05}$  \\ 
SN 1994I & Ic & 7.8 & $8.0^{+1.0}_{-1.2}$  & $-16.77^{+0.15}_{-0.15}$ & $0.52^{+0.24}_{-0.24}$ & $0.039^{+0.006}_{-0.005}$  \\ 
SN 1996cb & IIb & 17.0 & $24.2^{+0.6}_{-0.6}$  & $-17.38^{+0.06}_{-0.06}$ & $5.22^{+0.17}_{-0.17}$ & $0.16^{+0.01}_{-0.01}$  \\ 
SN 1997ef & Ic BL & 52.2 & $7.5^{+7.5}_{-7.5}$  & $-17.96^{+0.86}_{-0.86}$ & $0.43^{+2.07}_{-0.38}$ & $0.11^{+0.13}_{-0.06}$  \\ 
SN 1998bw & Ic BL & 37.9 & $11.3^{+2.8}_{-3.8}$  & $-19.50^{+0.23}_{-0.23}$ & $1.42^{+0.78}_{-0.99}$ & $0.66^{+0.16}_{-0.13}$  \\ 
SN 1999ex & Ic & 41.0 & $14.0^{+2.2}_{-2.2}$  & $-16.81^{+0.08}_{-0.08}$ & $2.20^{+0.67}_{-0.66}$ & $0.065^{+0.005}_{-0.005}$  \\ 
SN 2002ap & Ic BL & 9.3 & $15.6^{+0.6}_{-0.6}$  & $-17.77^{+0.17}_{-0.17}$ & $2.69^{+0.18}_{-0.18}$ & $0.17^{+0.03}_{-0.03}$  \\ 
SN 2003bg & Ic & 20.2 & $29.9^{+0.2}_{-0.4}$  & $-17.47^{+0.24}_{-0.24}$ & $6.80^{+0.06}_{-0.12}$ & $0.19^{+0.05}_{-0.04}$  \\ 
SN 2003id & Ic & 32.6 & $11.3^{+5.4}_{-8.1}$  & $-16.21^{+0.41}_{-0.41}$ & $1.42^{+1.57}_{-1.36}$ & $0.032^{+0.015}_{-0.010}$  \\ 
SN 2004aw & Ic & 71.7 & $18.7^{+2.4}_{-2.6}$  & $-17.60^{+0.14}_{-0.14}$ & $3.65^{+0.65}_{-0.78}$ & $0.17^{+0.02}_{-0.02}$  \\ 
SN 2004dk & Ib & 20.0 & $25.1^{+1.6}_{-11.1}$  & $-17.08^{+0.16}_{-0.16}$ & $5.44^{+0.45}_{-3.24}$ & $0.12^{+0.02}_{-0.02}$  \\ 
SN 2004dn & Ic & 56.3 & $16.6^{+3.4}_{-1.4}$  & $-16.98^{+0.22}_{-0.22}$ & $2.99^{+1.02}_{-0.43}$ & $0.086^{+0.020}_{-0.016}$  \\ 
SN 2004ex & IIb & 80.8 & $14.5^{+2.6}_{-3.1}$  & $-17.18^{+0.14}_{-0.14}$ & $2.38^{+0.73}_{-0.96}$ & $0.095^{+0.013}_{-0.011}$  \\ 
SN 2004fe & Ic & 80.8 & $12.6^{+2.6}_{-2.9}$  & $-17.55^{+0.14}_{-0.14}$ & $1.78^{+0.79}_{-0.86}$ & $0.12^{+0.02}_{-0.01}$  \\ 
SN 2004ff & IIb & 104.0 & $13.7^{+2.2}_{-2.2}$  & $-17.51^{+0.11}_{-0.11}$ & $2.08^{+0.67}_{-0.60}$ & $0.12^{+0.01}_{-0.01}$  \\ 
SN 2004gq & Ib & 28.2 & $15.5^{+2.2}_{-2.4}$  & $-17.17^{+0.10}_{-0.10}$ & $2.63^{+0.67}_{-0.67}$ & $0.098^{+0.009}_{-0.008}$  \\ 
SN 2004gt & Ic & 18.1 & $19.1^{+4.2}_{-5.4}$  & $-16.27^{+0.14}_{-0.14}$ & $3.71^{+1.22}_{-1.57}$ & $0.049^{+0.007}_{-0.006}$  \\ 
SN 2004gv & Ib & 90.0 & $17.8^{+2.9}_{-2.5}$  & $-17.69^{+0.09}_{-0.09}$ & $3.35^{+0.89}_{-0.79}$ & $0.17^{+0.02}_{-0.01}$  \\ 
SN 2005aw & Ic & 42.3 & $15.1^{+3.6}_{-5.3}$  & $-17.47^{+0.23}_{-0.23}$ & $2.50^{+1.09}_{-1.53}$ & $0.13^{+0.03}_{-0.02}$  \\ 
SN 2005az & Ic & 33.0 & $17.6^{+6.5}_{-7.7}$  & $-16.54^{+0.30}_{-0.30}$ & $3.29^{+1.92}_{-2.27}$ & $0.060^{+0.019}_{-0.014}$  \\ 
SN 2005bf & Ib & 84.9 & $13.7^{+1.2}_{-1.8}$  & $-18.53^{+0.09}_{-0.09}$ & $2.14^{+0.37}_{-0.54}$ & $0.31^{+0.03}_{-0.03}$  \\ 
SN 2005bj & Ic & 99.0 & $16.4^{+3.5}_{-4.5}$  & $-17.56^{+0.20}_{-0.20}$ & $2.93^{+1.08}_{-1.33}$ & $0.15^{+0.03}_{-0.02}$  \\ 
SN 2005em & IIb & 113.5 & $13.3^{+2.0}_{-2.0}$  & $-17.75^{+0.09}_{-0.09}$ & $1.96^{+0.61}_{-0.54}$ & $0.15^{+0.01}_{-0.01}$  \\ 
SN 2005hg & Ib & 94.4 & $19.2^{+10.4}_{-10.2}$  & $-17.65^{+0.22}_{-0.22}$ & $3.77^{+2.97}_{-3.01}$ & $0.18^{+0.04}_{-0.03}$  \\ 
SN 2005hl & Ib & 104.4 & $17.0^{+5.9}_{-6.6}$  & $-17.14^{+0.23}_{-0.23}$ & $3.11^{+1.70}_{-1.98}$ & $0.10^{+0.02}_{-0.02}$  \\ 
SN 2005hm & Ib & 158.0 & $11.2^{+2.9}_{-2.1}$  & $-17.24^{+0.05}_{-0.05}$ & $1.36^{+0.90}_{-0.60}$ & $0.081^{+0.004}_{-0.004}$  \\ 
SN 2005kl & Ic & 22.2 & $20.1^{+11.6}_{-15.1}$  & $-16.85^{+0.37}_{-0.37}$ & $4.07^{+3.26}_{-4.01}$ & $0.087^{+0.035}_{-0.025}$  \\ 
SN 2005kr & Ic BL & 650.1 & $11.3^{+2.5}_{-5.6}$  & $-19.29^{+0.16}_{-0.16}$ & $1.36^{+0.78}_{-1.26}$ & $0.54^{+0.09}_{-0.07}$  \\ 
SN 2006T & IIb & 31.0 & $16.1^{+3.0}_{-4.0}$  & $-16.99^{+0.17}_{-0.17}$ & $2.81^{+0.90}_{-1.15}$ & $0.085^{+0.014}_{-0.012}$  \\ 
SN 2006ep & Ib & 63.0 & $14.6^{+2.0}_{-2.0}$  & $-16.85^{+0.11}_{-0.11}$ & $2.38^{+0.61}_{-0.61}$ & $0.070^{+0.007}_{-0.007}$  \\ 
SN 2006fo & Ib & 93.1 & $21.4^{+5.7}_{-7.2}$  & $-17.48^{+0.14}_{-0.14}$ & $4.42^{+1.59}_{-2.16}$ & $0.16^{+0.02}_{-0.02}$  \\ 
SN 2006jo & Ib & 360.0 & $10.2^{+2.0}_{-5.0}$  & $-18.00^{+0.22}_{-0.22}$ & $1.08^{+0.58}_{-1.01}$ & $0.15^{+0.03}_{-0.03}$  \\ 
SN 2006lc & Ib & 59.0 & $12.6^{+2.2}_{-2.6}$  & $-16.60^{+0.10}_{-0.10}$ & $1.78^{+0.67}_{-0.75}$ & $0.050^{+0.005}_{-0.004}$  \\ 
SN 2006nx & Ic & 603.9 & $11.6^{+4.8}_{-4.8}$  & $-19.02^{+0.22}_{-0.22}$ & $1.48^{+1.45}_{-1.20}$ & $0.43^{+0.10}_{-0.08}$  \\ 
SN 2007C & Ib & 29.0 & $12.7^{+2.6}_{-2.6}$  & $-16.83^{+0.12}_{-0.12}$ & $1.84^{+0.79}_{-0.76}$ & $0.062^{+0.007}_{-0.006}$  \\ 
SN 2007Y & Ib & 19.3 & $12.7^{+2.0}_{-2.2}$  & $-16.25^{+0.15}_{-0.15}$ & $1.84^{+0.61}_{-0.64}$ & $0.037^{+0.005}_{-0.005}$  \\ 
SN 2007ay & IIb & 67.1 & $8.9^{+3.3}_{-3.0}$  & $-17.38^{+0.27}_{-0.27}$ & $0.71^{+0.95}_{-0.56}$ & $0.075^{+0.021}_{-0.016}$  \\ 
SN 2007cl & Ic & 99.9 & $14.7^{+5.7}_{-8.8}$  & $-17.39^{+0.31}_{-0.31}$ & $2.38^{+1.74}_{-2.25}$ & $0.12^{+0.04}_{-0.03}$  \\ 
SN 2007gr & Ic & 10.0 & $11.4^{+2.2}_{-1.8}$  & $-17.33^{+0.19}_{-0.19}$ & $1.42^{+0.66}_{-0.50}$ & $0.089^{+0.017}_{-0.014}$  \\ 
SN 2007hn & Ib/c & 140.0 & $21.2^{+6.4}_{-4.3}$  & $-17.68^{+0.19}_{-0.19}$ & $4.36^{+1.82}_{-1.31}$ & $0.19^{+0.04}_{-0.03}$  \\ 
SN 2007kj & Ib & 80.3 & $9.4^{+2.0}_{-1.6}$  & $-17.25^{+0.12}_{-0.12}$ & $0.87^{+0.56}_{-0.39}$ & $0.070^{+0.008}_{-0.007}$  \\ 
SN 2007ru & Ic BL & 69.4 & $14.8^{+2.8}_{-2.6}$  & $-19.06^{+0.26}_{-0.26}$ & $2.44^{+0.85}_{-0.79}$ & $0.54^{+0.14}_{-0.11}$  \\ 
SN 2007uy & Ib & 27.0 & $14.9^{+7.2}_{-10.1}$  & $-16.53^{+0.30}_{-0.30}$ & $2.50^{+2.08}_{-2.45}$ & $0.053^{+0.017}_{-0.013}$  \\ 
SN 2008D & Ib & 27.0 & $18.1^{+3.4}_{-4.2}$  & $-15.42^{+0.10}_{-0.10}$ & $3.41^{+1.00}_{-1.21}$ & $0.022^{+0.002}_{-0.002}$  \\ 
SN 2008aq & IIb & 32.8 & $13.5^{+3.0}_{-3.4}$  & $-16.79^{+0.17}_{-0.17}$ & $2.02^{+0.91}_{-0.94}$ & $0.062^{+0.010}_{-0.009}$  \\ 
SN 2008ax & IIb & 6.5 & $21.0^{+3.0}_{-3.4}$  & $-15.68^{+0.13}_{-0.13}$ & $4.30^{+0.86}_{-1.01}$ & $0.030^{+0.004}_{-0.003}$  \\ 
SN 2008bo & IIb & 21.0 & $10.3^{+3.2}_{-2.8}$  & $-15.64^{+0.18}_{-0.18}$ & $1.14^{+0.94}_{-0.70}$ & $0.018^{+0.003}_{-0.003}$  \\ 
SN 2009bb & Ic & 46.4 & $16.0^{+1.8}_{-1.8}$  & $-17.80^{+0.11}_{-0.11}$ & $2.81^{+0.54}_{-0.55}$ & $0.18^{+0.02}_{-0.02}$  \\ 
SN 2009dp & Ic & 104.5 & $20.3^{+4.9}_{-4.9}$  & $-17.53^{+0.24}_{-0.24}$ & $4.13^{+1.37}_{-1.50}$ & $0.16^{+0.04}_{-0.03}$  \\ 
SN 2009iz & Ib & 62.6 & $21.5^{+7.7}_{-6.7}$  & $-17.24^{+0.15}_{-0.15}$ & $4.42^{+2.21}_{-1.97}$ & $0.13^{+0.02}_{-0.02}$  \\ 
SN 2009jf & Ib & 31.0 & $21.2^{+6.5}_{-7.5}$  & $-17.81^{+0.32}_{-0.32}$ & $4.36^{+1.87}_{-2.28}$ & $0.22^{+0.08}_{-0.06}$  \\ 
SN 2009mg & IIb & 37.0 & $9.7^{+9.7}_{-9.7}$  & $-18.15^{+0.60}_{-0.60}$ & $0.97^{+2.86}_{-0.91}$ & $0.17^{+0.12}_{-0.07}$  \\ 
PTF09cnd & SLSN-I & 1349.8 & $19.1^{+4.8}_{-3.5}$  & $-23.38^{+0.08}_{-0.08}$ & $3.71^{+1.39}_{-1.03}$ & $34.12^{+2.56}_{-2.38}$  \\ 
SN 2010md & SLSN-I & 468.8 & $18.6^{+3.5}_{-4.2}$  & $-21.26^{+0.20}_{-0.20}$ & $3.59^{+1.00}_{-1.27}$ & $4.79^{+0.99}_{-0.82}$  \\ 
PS1-10awh & SLSN-I & 6031.0 & $15.3^{+4.2}_{-4.2}$  & $-23.04^{+0.19}_{-0.19}$ & $2.57^{+1.27}_{-1.20}$ & $21.52^{+4.02}_{-3.39}$  \\ 
PS1-10ky & SLSN-I & 6425.0 & $10.4^{+3.6}_{-4.8}$  & $-23.04^{+0.24}_{-0.24}$ & $1.14^{+1.06}_{-1.03}$ & $15.88^{+3.84}_{-3.09}$  \\ 
SN 2011bm & Ic & 99.0 & $40.7^{+11.5}_{-13.5}$  & $-18.60^{+0.15}_{-0.15}$ & $10.40^{+13.27}_{-4.34}$ & $0.67^{+0.10}_{-0.08}$  \\ 
SN 2011dh & IIb & 8.0 & $18.0^{+2.6}_{-2.8}$  & $-17.22^{+0.09}_{-0.09}$ & $3.41^{+0.77}_{-0.85}$ & $0.11^{+0.01}_{-0.01}$  \\ 
SN 2011hs & IIb & 30.3 & $9.7^{+1.8}_{-2.0}$  & $-16.89^{+0.14}_{-0.14}$ & $0.97^{+0.51}_{-0.50}$ & $0.052^{+0.007}_{-0.006}$  \\ 
SN 2011ke & SLSN-I & 697.8 & $7.4^{+0.2}_{-0.5}$  & $-22.72^{+0.26}_{-0.26}$ & $0.39^{+0.04}_{-0.11}$ & $8.69^{+2.31}_{-1.82}$  \\ 
SN 2011kg & SLSN-I & 968.7 & $21.5^{+1.5}_{-1.5}$  & $-21.18^{+0.09}_{-0.09}$ & $4.42^{+0.46}_{-0.41}$ & $4.87^{+0.44}_{-0.41}$  \\
PS1-11ap & SLSN-I & 3087.0 & $33.7^{+5.1}_{-3.1}$  & $-22.26^{+0.06}_{-0.06}$ & $7.98^{+1.63}_{-0.95}$ & $16.97^{+1.00}_{-0.94}$  \\ 
SN 2012P & Ib/c & 27.4 & $21.1^{+6.8}_{-7.4}$  & $-16.31^{+0.11}_{-0.11}$ & $4.36^{+1.87}_{-2.22}$ & $0.054^{+0.006}_{-0.005}$  \\ 
SN 2012ap & Ic BL & 54.7 & $13.8^{+1.0}_{-1.0}$  & $-17.71^{+0.09}_{-0.09}$ & $2.14^{+0.30}_{-0.30}$ & $0.15^{+0.01}_{-0.01}$  \\ 
SN 2012au & Ib & 23.6 & $8.8^{+4.8}_{-3.2}$  & $-18.60^{+0.20}_{-0.20}$ & $0.71^{+1.37}_{-0.60}$ & $0.23^{+0.05}_{-0.04}$  \\ 
SN 2012bz & Ic & 1498.0 & $10.0^{+4.5}_{-5.1}$  & $-19.48^{+0.36}_{-0.36}$ & $1.03^{+1.30}_{-0.97}$ & $0.58^{+0.23}_{-0.16}$  \\ 
SN 2012hn & Ic & 33.8 & $10.1^{+4.4}_{-3.4}$  & $-15.89^{+0.27}_{-0.27}$ & $1.08^{+1.24}_{-0.80}$ & $0.022^{+0.006}_{-0.005}$  \\ 
PS1-12bqf & SLSN-I & 3073.0 & $28.3^{+15.0}_{-6.3}$  & $-21.12^{+0.08}_{-0.08}$ & $6.34^{+5.24}_{-1.76}$ & $5.36^{+0.41}_{-0.38}$  \\ 
PTF12dam & SLSN-I & 512.5 & $26.5^{+1.4}_{-1.4}$  & $-22.82^{+0.09}_{-0.09}$ & $5.89^{+0.40}_{-0.39}$ & $24.90^{+2.21}_{-2.03}$  \\ 
SN 2013ak & IIb & 7.0 & $18.9^{+3.2}_{-9.2}$  & $-16.32^{+0.19}_{-0.19}$ & $3.71^{+0.93}_{-2.74}$ & $0.051^{+0.010}_{-0.008}$  \\ 
SN 2013hy & SLSN-I & 4103.0 & $11.8^{+7.1}_{-2.2}$  & $-21.20^{+0.11}_{-0.11}$ & $1.54^{+2.11}_{-0.62}$ & $3.24^{+0.35}_{-0.32}$  \\ 
iPTF13ajg & SLSN-I & 4690.0 & $18.3^{+1.6}_{-2.2}$  & $-23.01^{+0.16}_{-0.16}$ & $3.47^{+0.48}_{-0.67}$ & $23.66^{+3.71}_{-3.21}$  \\ 
iPTF13bvn & Ib & 22.5 & $9.4^{+1.8}_{-1.6}$  & $-16.86^{+0.09}_{-0.09}$ & $0.87^{+0.50}_{-0.39}$ & $0.049^{+0.004}_{-0.004}$  \\
SN 2014L & Ic & 35.8 & $13.3^{+2.6}_{-2.8}$  & $-18.54^{+0.16}_{-0.16}$ & $1.96^{+0.79}_{-0.76}$ & $0.31^{+0.05}_{-0.04}$  \\ 
SN 2014ad & Ic BL & 25.3 & $8.9^{+4.8}_{-5.0}$  & $-18.24^{+0.39}_{-0.39}$ & $0.76^{+1.38}_{-0.70}$ & $0.17^{+0.07}_{-0.05}$  \\ 
SN 2015ap & Ib & 50.8 & $15.2^{+5.1}_{-5.1}$  & $-18.61^{+0.20}_{-0.20}$ & $2.57^{+1.56}_{-1.54}$ & $0.36^{+0.07}_{-0.06}$  \\ 
SN 2015bn & SLSN-I & 544.8 & $17.2^{+1.6}_{-1.8}$  & $-22.64^{+0.06}_{-0.06}$ & $3.17^{+0.48}_{-0.55}$ & $16.22^{+1.00}_{-0.94}$  \\ 
SN 2016coi & Ic BL & 16.2 & $8.0^{+4.6}_{-4.6}$  & $-17.51^{+0.21}_{-0.21}$ & $0.52^{+1.26}_{-0.46}$ & $0.077^{+0.016}_{-0.014}$  \\ 
SN 2016eay & SLSN-I & 481.9 & $22.2^{+2.4}_{-2.7}$  & $-22.56^{+0.12}_{-0.12}$ & $4.65^{+0.68}_{-0.82}$ & $17.71^{+2.00}_{-1.80}$  \\ 
PS16yj & SLSN-I & 1148.0 & $12.3^{+23.2}_{-4.6}$  & $-21.02^{+0.13}_{-0.13}$ & $1.66^{+6.82}_{-1.22}$ & $2.84^{+0.35}_{-0.31}$  \\ 
SN 2017ein & Ic & 17.0 & $13.2^{+5.0}_{-5.4}$  & $-16.38^{+0.15}_{-0.15}$ & $1.96^{+1.52}_{-1.48}$ & $0.042^{+0.006}_{-0.005}$  \\ 
DES17C1ffz & Ib/c & 427.5 & $16.5^{+6.4}_{-6.6}$  & $-17.67^{+0.25}_{-0.25}$ & $2.99^{+1.88}_{-1.97}$ & $0.16^{+0.04}_{-0.03}$  \\ 
\enddata
\label{tab:summary}

\end{deluxetable*}

\end{document}